\documentclass[usenatbib]{mnras}
\usepackage{aas_macros}
\usepackage{ae,aecompl}
\usepackage[dvipsnames]{xcolor}
\hypersetup{colorlinks=true,allcolors=teal}
\usepackage[english]{babel}
\usepackage{amsmath}
\usepackage{amssymb}
\usepackage{graphicx}

\newcommand{\Mh}{\ensuremath{h^{-1}M_{\odot}}}

\newcommand{\Mpch}{\ensuremath{h^{-1}{\rm Mpc}}}

\newcommand{\der}{\ensuremath{{\rm d}}}

\newcommand{\erf}[1]{\ensuremath{{\rm erf}\left(#1\right)}}

\newcommand{\Nscript}{\ensuremath{\mathcal{N}_{\rm sat}}}

\newcommand{\fcen}{\ensuremath{f_{\rm cen}}}
\newcommand{\Ns}{\ensuremath{\bar N_{\rm sat}}}

\newcommand{\eqn}[1]{equation~\eqref{#1}}

\newcommand{\be}{\begin{equation}}
\newcommand{\ee}{\end{equation}}
\newcommand{\ph}[1]{\phantom{#1}}

\title[Global HOD analysis]{Global analysis of luminosity- and colour-dependent galaxy clustering in the Sloan Digital Sky Survey}
\author[Paul, Pahwa \& Paranjape]{Niladri Paul, Isha Pahwa
\& Aseem Paranjape\thanks{E-mail: npaul, ipahwa \& aseem@iucaa.in}
\\
Inter-University Centre for Astronomy and Astrophysics, Ganeshkhind, Post Bag 4, Pune 411007, India.}
\date{draft}
\begin{document}
\label{firstpage}
\pagerange{\pageref{firstpage}--\pageref{lastpage}}
\maketitle
\begin{abstract}
We present a Halo Occupation Distribution (HOD) analysis of the luminosity- and colour-dependent galaxy clustering in the Sloan Digital Sky Survey. 
A novelty of our technique is that it uses a combination of clustering measurements in luminosity bins to perform a global likelihood analysis, simultaneously constraining the HOD parameters for a range of luminosity thresholds. 
We present simple, smooth fitting functions which accurately describe the resulting luminosity dependence of the best-fit HOD parameters. 
To minimise systematic halo modelling effects, we use theoretical halo 2-point correlation functions directly measured and tabulated from a suite of $N$-body simulations spanning a large enough dynamic range in halo mass and spatial separation. 
Thus, our modelling correctly accounts for non-linear and scale-dependent halo bias as well as any departure of halo profiles from universality, and we additionally account for halo exclusion using the hard sphere approximation.  
Using colour-dependent clustering information, we constrain the satellite galaxy red fraction in a model-independent manner which does not rely on any group-finding algorithm. 
We find that the resulting luminosity dependence of the satellite red fraction is significantly shallower than corresponding measurements from galaxy group catalogues, and we provide a simple fitting function to describe this dependence. 
Our fitting functions are readily usable in generating low-redshift mock galaxy catalogues, and we discuss some potentially interesting applications as well as possible extensions of our technique. 
\end{abstract}
\section{Introduction} \label{sec:intro}
\noindent
Understanding the physical mechanisms at play in the formation and evolution of galaxies, and their connection to the underlying dark matter distribution and cosmology, is a problem of great interest \citep{White_Rees_1978, Mo_White_Bosch_book_2010, Somerville_Dave_2015}. To understand the large-scale distribution of the baryons, one needs to run computationally expensive full hydrodynamic large-scale simulations \citep{Vogelsberger_et_al_2014, Genel_et_al_2014}. There are different semi-numerical techniques to overcome these issues. One of the popular techniques is semi-analytical modelling of galaxy formation \citep{Lacey_et_al_2016, Zoldan_et_al_2017, Gonzalez_Perez_et_al_2018} where people use simplified mathematical formulae to understand the baryonic processes affecting galaxy evolution, happening inside halos, e.g. star formation, supernovae feedback, AGN feedback, gas cooling, tidal stripping etc. 

Alternatively, one can statistically model the mapping between dark matter and galaxies assuming the halo model \citep{Cooray_Sheth_2002} using the halo occupation distribution (HOD) \citep{Berlind_Weinberg_2002} or conditional luminosity function (CLF) approaches \citep{Yang_et_al_2008, Cacciato_et_al_2013}.   
In the HOD formalism which we adopt in this work, one prescribes a statistical routine about how to populate the halos with galaxies depending on halo and galaxy properties \citep{Seljak_et_al_2000, Scoccimarro_et_al_2001}.  It has been seen that it is necessary to split the galaxy population into centrals and satellites to describe both the correlation and abundance data of the galaxies accurately \citep{Berlind_et_al_2003, Yang_et_al_2003, Zheng_et_al_2004, Vale_Ostriker_2004, Zehavi_et_al_2005}. 

Starting from the simple HOD model which depends only on the mass of the halos, there have been several modifications to include `beyond halo mass' effects \citep{Wechsler_et_al_2006, hearin+15a, Tinker_et_al_2017, Ross_Brunner_2009,pkhp15, pp17b, Xu_et_al_2018, Lange_et_al_2018}, `velocity bias' effects \citep{Bosch_et_al_2005, behroozi13-rockstar, Reid_et_al_2014, Guo_et_al_2015}, the deviation of satellite density profiles from the dark matter halo profile \citep{Yang_et_al_2005, Chen_et_al_2008, More_et_al_2009, Guo_quan_et_al_2012, Watson_et_al_2012} etc. These, however, lead to relatively minor contributions in the luminosity- and colour-dependent clustering when galaxies are not explicitly classified as being centrals or satellites. We will therefore ignore these effects in this work and focus on the simplest, `halo mass only' flavour of HOD models. Other effects such as scale-dependent halo bias \citep{Tinker_et_al_2005}, halo exclusion \citep{Tinker_et_al_2012, Leauthaud_et_al_2011} and the adopted calibration of the halo concentration-mass relation \citep{Wechsler_et_al_2002,Zhao_et_al_2003, Lu_et_al_2006, Ludlow_et_al_2013, Ludlow_et_al_2014} can lead to systematic biases of order $\gtrsim10\%$ in HOD constraints and are therefore important to be modelled accurately \citep{V_Bosch_et_al_2013}. 

One way to take into account of all these effects consistently in one go is to directly use the measurements of halo correlation functions from $N$-body simulations \citep{zg16}. This way of accurate calibrations of halo model parameters has applications in understanding the redshift evolution of galaxy population inside halos \citep{Zheng_et_al_2007, White_et_al_2007, Wake_et_al_2008, Wake_et_al_2011, Abbas_et_al_2010, Coupon_et_al_2012, de_la_torre_et_al_2013, Guo_et_al_2014, Manera_et_al_2015, Skibba_et_al_2015, contreras+17}, reconstructing the initial conditions from large csale galaxy surveys etc. \citep{Nusser_et_al_1992,  Crocce_Scoccimarro_2006}. This is the approach we will adopt in this paper. 

Additionally, in this work, we will describe a novel approach of global HOD fitting which allows us to combine clustering measurements for a wide range of galaxy luminosities in a statistically consistent manner. The resulting HOD parameters turn out to vary smoothly with luminosity threshold, and we fit them using simple, smooth functions. We will also use colour-dependent clustering information to constrain the red fraction of satellites in the data volume.

This article is organised as follows. In section~\ref{sec:sim_and_data}, we describe our simulation suite and our measurements of the halo 2-point correlation function (2pcf) from it. We also describe the observational data set we use, along with a comparison of the theoretical errors in our simulation-based model with the corresponding observational errors. In section~\ref{sec:global_HOD}, we describe our global likelihood analysis of the SDSS projected clustering and abundances, along with a discussion of the data covariance matrices. In section~\ref{sec:results}, we present the results of using luminosity- and colour-dependent clustering to constrain HOD parameters and the satellite red fraction as a function of luminosity, and present our fitting functions for these quantities. We conclude in section~\ref{sec:conclude}. The Appendices describe some technical details of results used in our analysis.

Throughout, we will adopt a flat $\Lambda$CDM cosmology  with total matter density parameter $\Omega_{\rm m} = 0.276$, baryonic matter density $\Omega_{\rm b} = 0.045$, Hubble constant $H_0 = 100 h \rm km s^{-1}Mpc^{-1}$ with $h = 0.7$, primordial scalar spectral index $n_{\rm s} = 0.961$ and root mean square linear fluctuations in spheres of radius $8 \Mpch$, $\sigma_8=0.811$. 

\section{Simulations and observational data} \label{sec:sim_and_data}
In this section, we describe our simulation suite and the observational data set we will use in this work. We also describe our tabulated measurements of the halo 2pcf which will serve as the basis of our theoretical model and demonstrate that the corresponding statistical uncertainties are substantially smaller than the errors in the observational measurements.

\subsection{Simulation details} \label{sec:sim_details}
To resolve a full dynamic range of halos ranging from $10^9 \Mh$ to $10^{15} \Mh$, one needs to run a box of size $1 h^{-1} {\rm Gpc}$ with approximately $3048^3$ particles. To minimise statistical variance, one would need to run several realisations of this high-resolution simulation which requires a huge computational budget. To overcome this issue, we instead combine three simulation boxes of size $150 \Mpch$, $300 \Mpch$ and $600 \Mpch$, each run with $1024^3$ particles. We will refer to these as the small, intermediate and large box in what follows. The dark matter particle mass in these boxes were $m_{\rm p}= 2.41 \times 10^8\Mh$, $1.93 \times 10^9 \Mh$ and $1.54 \times 10^{10} \Mh$, respectively. The force resolution $\epsilon$ in each case was set equal to $1/30$ of the mean comoving inter-particle spacing, leading to $\epsilon = 4.9, 9.8, 19.5 h^{-1} {\rm kpc}$ for the $150, 300, 600\Mpch$ boxes, respectively.

All simulations used cold dark matter only and were performed using the tree-PM $N$-body code \textsc{gadget-2}\footnote{\href{http://www.mpa-garching.mpg.de/gadget/}{http://www.mpa-garching.mpg.de/gadget/}} \citep{Springel_2005}. Initial conditions were set at a starting redshift $z_{\rm in}=99, 49, 99$ in case of $150, 300 $ and $600 \Mpch$ box respectively using the code \textsc{music}\footnote{\href{https://www-n.oca.eu/ohahn/MUSIC/}{https://www-n.oca.eu/ohahn/MUSIC/}} \citep{hahn11-music} with 2nd order Lagrangian perturbation theory. By changing the random number seed for the initial conditions, we generated $3$ realisations of the large box and $10$ realisations of the intermediate box while using a single realisation of the small box.
Halos were identified using the code \textsc{rockstar}\footnote{\href{http://code.google.com/p/rockstar/}{http://code.google.com/p/rockstar/}} \citep{behroozi13-rockstar} which performs a Friends-of-Friends (FoF) algorithm in 6-dimensional phase space. Throughout, we will use $m$ to denote the mass of the halo in a radius $R_{\rm 200b}$ that encloses a dark matter density of $200$ times the mean density of the universe. Wherever needed, we will use $R_{\rm 200b}$ as the halo radius. The simulations and analysis were performed on the Perseus cluster at IUCAA.\footnote{\href{http://hpc.iucaa.in}{http://hpc.iucaa.in}}

\subsection{Calculating halo statistics from simulations} \label{sec:sim_stats}
Our $N$-body simulations allow us to accurately and self-consistently account for scale-dependent bias, halo exclusion and the nature of dark matter profiles in halos. In order to do this, we have combined measurements of the halo 2pcf over a wide range of halo mass and spatial separation, as we describe next.

The single realisation of our small box allows us to reliably obtain the mass function and the 1-halo term of the correlation function in the mass range $10.6 < \log [m/ \Mh] < 11.3$ (see Appendix~\ref{sec:2PCF_calculation_formulae} for the relevant details of halo model formalism). The $10$ realisations of the intermediate box similarly give us estimates of halo mass function for $11.0 < \log[m/\Mh] < 15.5$ and 1-halo information for $11.3 <\log[m/\Mh]< 14.7$. Additionally, these realisations also allow us to measure the 2-halo term of the correlation function in the entire halo mass range $10.6 < \log [m/\Mh] < 15.5$ which we use in this work.  The $3$  realisations of the large box provide us reliable estimates of the mass function and 2-halo correlation function measurements for $13.0 < \log [m/\Mh] < 15.5$, as well as 1-halo information for $14.0 < \log [m/\Mh] < 15.5$.  

We follow the simulation-based method developed in \cite{zg16} to compute the galaxy 2pcf. In this approach, we place one galaxy at the halo centre and choose halo particles as tracers of the satellite galaxies. We create tables for halo properties, including halo number density (i.e., halo mass function) and real-space 2pcf of 1-halo and 2-halo terms. 

Each correlation function consists of five terms - 1-halo cen-sat, 1-halo sat-sat, 2-halo cen-cen, 2-halo cen-sat and 2-halo sat-sat. For the 1-halo terms, we use all the particles in the halo. However, in order to reduce the time taken to compute the 2-halo terms, we randomly choose $50$ particles in the halo if the halo has more than $50$ particles and suitably normalise the resulting measurements. We have checked that using $50$ particles per halo gives us sufficient precision in the resulting measurements (see also section~\ref{sec:sim_errors} below). We have additionally incorporated hard-sphere exclusion by only counting halo pairs whose separation is larger than the sum of their virial radii. To generate the tables, we divide the halos into logarithmic mass bins of width $0.1$dex each. 

\begin{figure}
\centering
\includegraphics[scale=0.5]{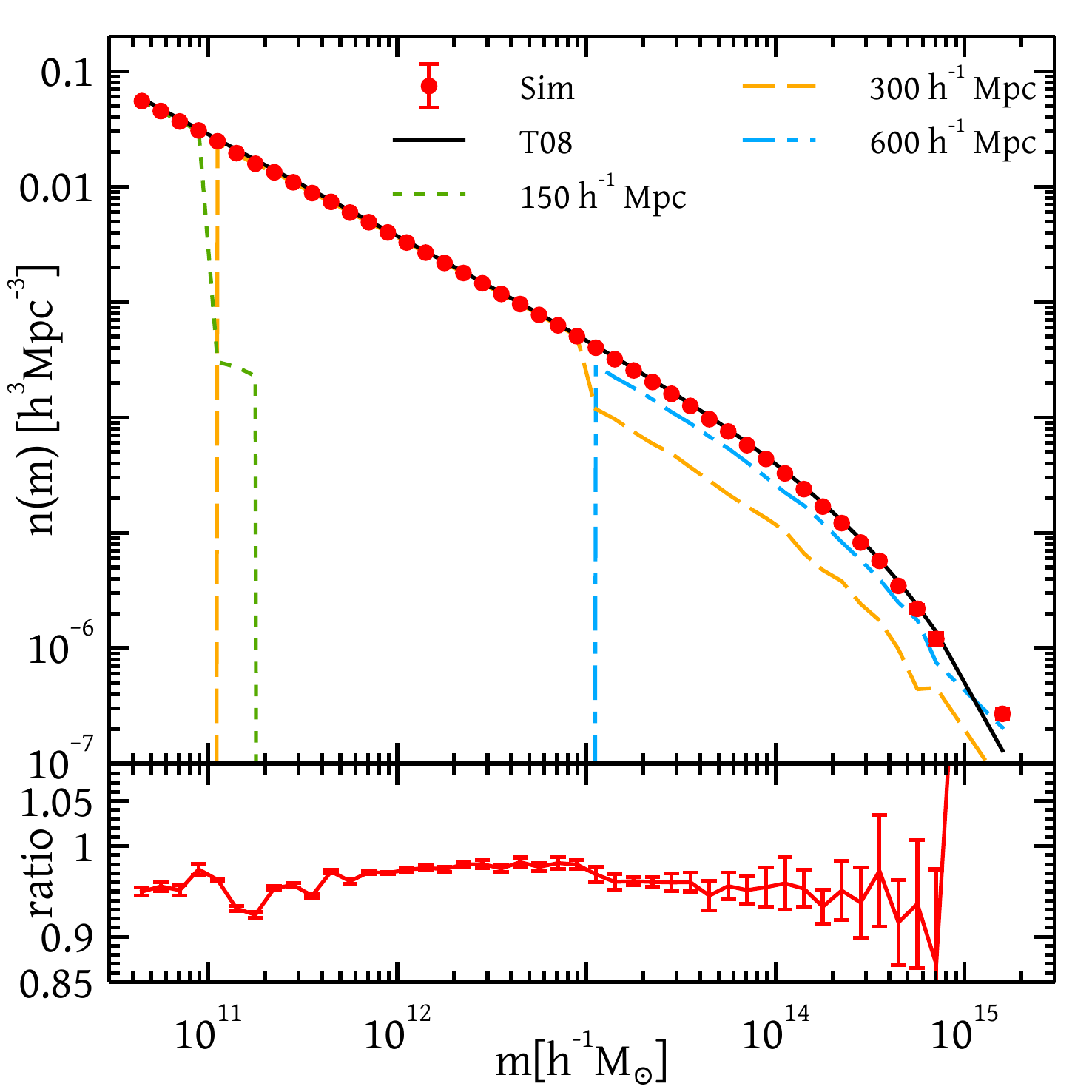} 
\caption{Halo mass function measured from our suite of simulations. The \emph{upper panel} shows the weighted average of measurements from all the realisations (red points with error bars), with individual weighted contributions from each box shown with different line styles and colours as indicated, compared with the fitting formula of \citet[][solid black line]{Tinker_2008_mass_function}. The \emph{lower panel} shows the ratio of the full measurements and the Tinker fit. See the main text and Appendix~\ref{sec:simulation_weight} for a description of our weighting procedure. }
\label{fig:hmf_from_sim}
\end{figure}
\begin{figure}
\centering
\includegraphics[scale=0.5]{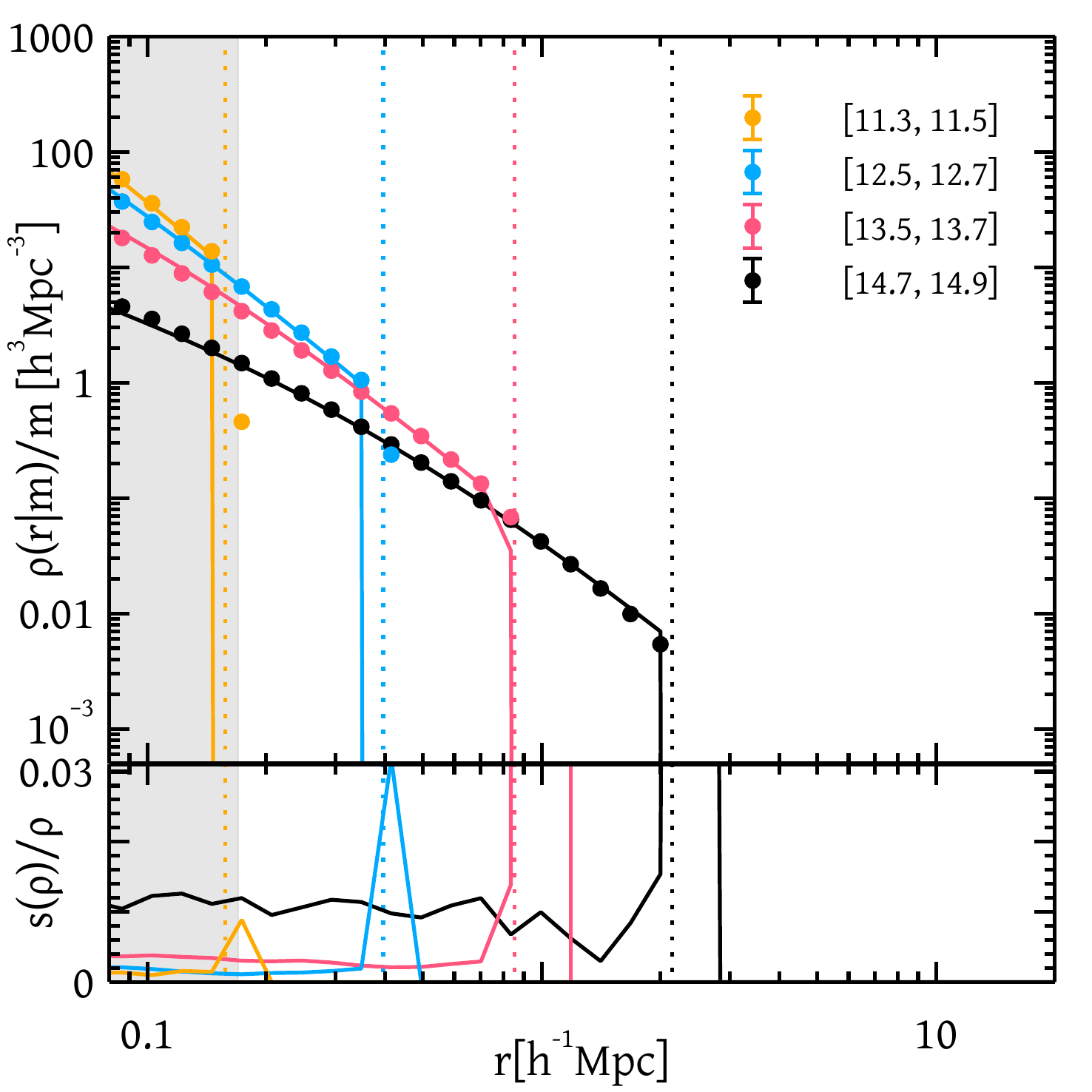}
\caption{
Halo density profiles $\rho(r|m)$ measured from our suite of simulations. The \emph{upper panel} shows weighted measurements of $\rho(r|m)/m$ from simulations in different halo mass bins (coloured circles with error bars), compared with the analytical NFW \citep{Navarro_1997} profiles computed with a median concentration-mass relation found from the simulations (solid curves). The vertical dotted lines of different colours denote virial radii ($R_{\rm 200b}$) for the corresponding halo mass bins. The \emph{lower panel} shows the relative statistical errors on these measurements, computed as described in the main text and Appendix~\ref{sec:simulation_weight}. In both panels, the grey shaded region denotes the range of separations which is excluded in our HOD routines.} 
\label{fig:rho_from_sim}
\end{figure}
\begin{figure}
\centering
\includegraphics[scale=0.5]{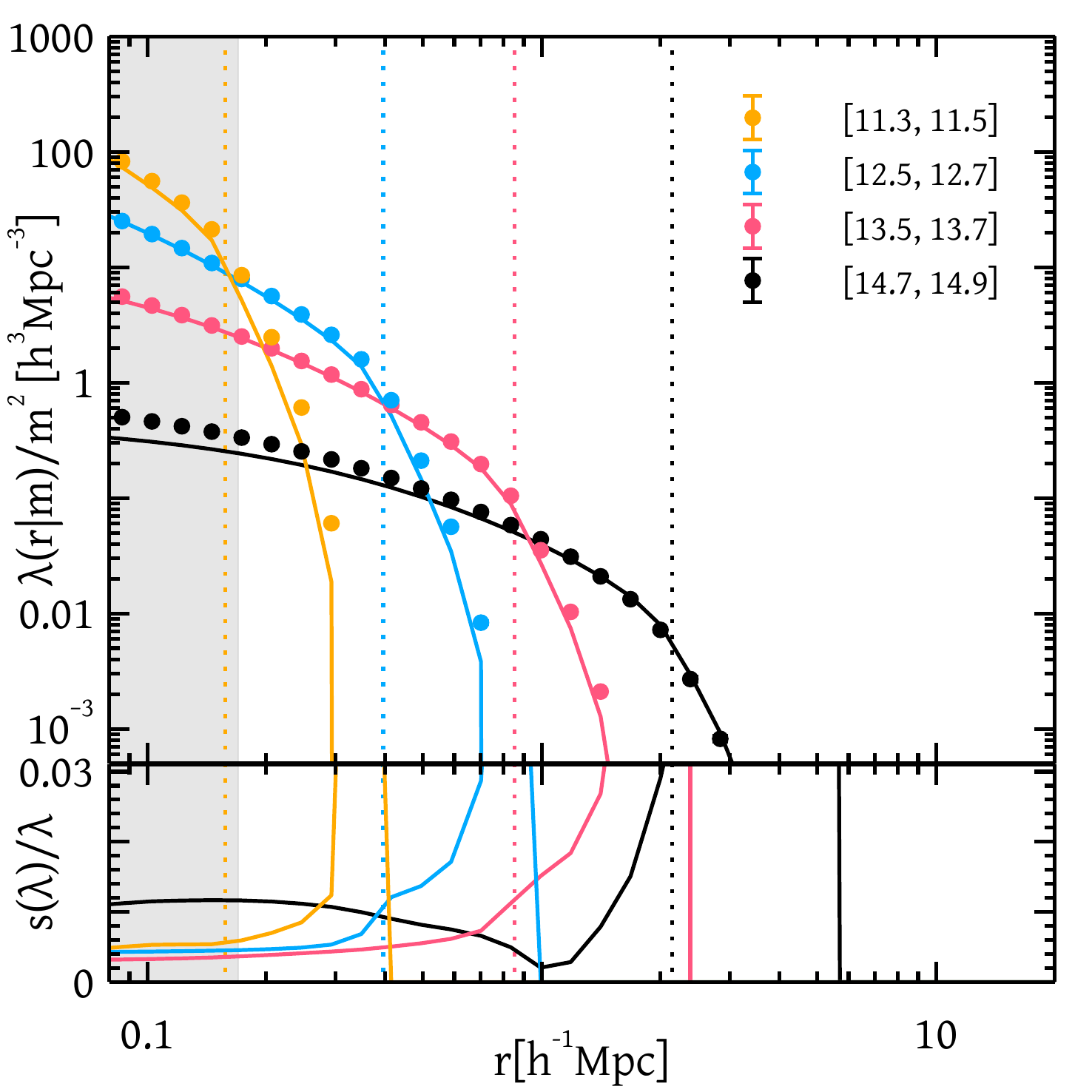} 
\caption{Similar to Figure~\ref{fig:rho_from_sim}, showing  $\lambda(r|m)/m^2$ where $\lambda(r|m)$ is the convolution of the halo density profile with itself. } 
\label{fig:lamda_from_sim}
\end{figure}
\begin{figure}
\includegraphics[scale=0.5]{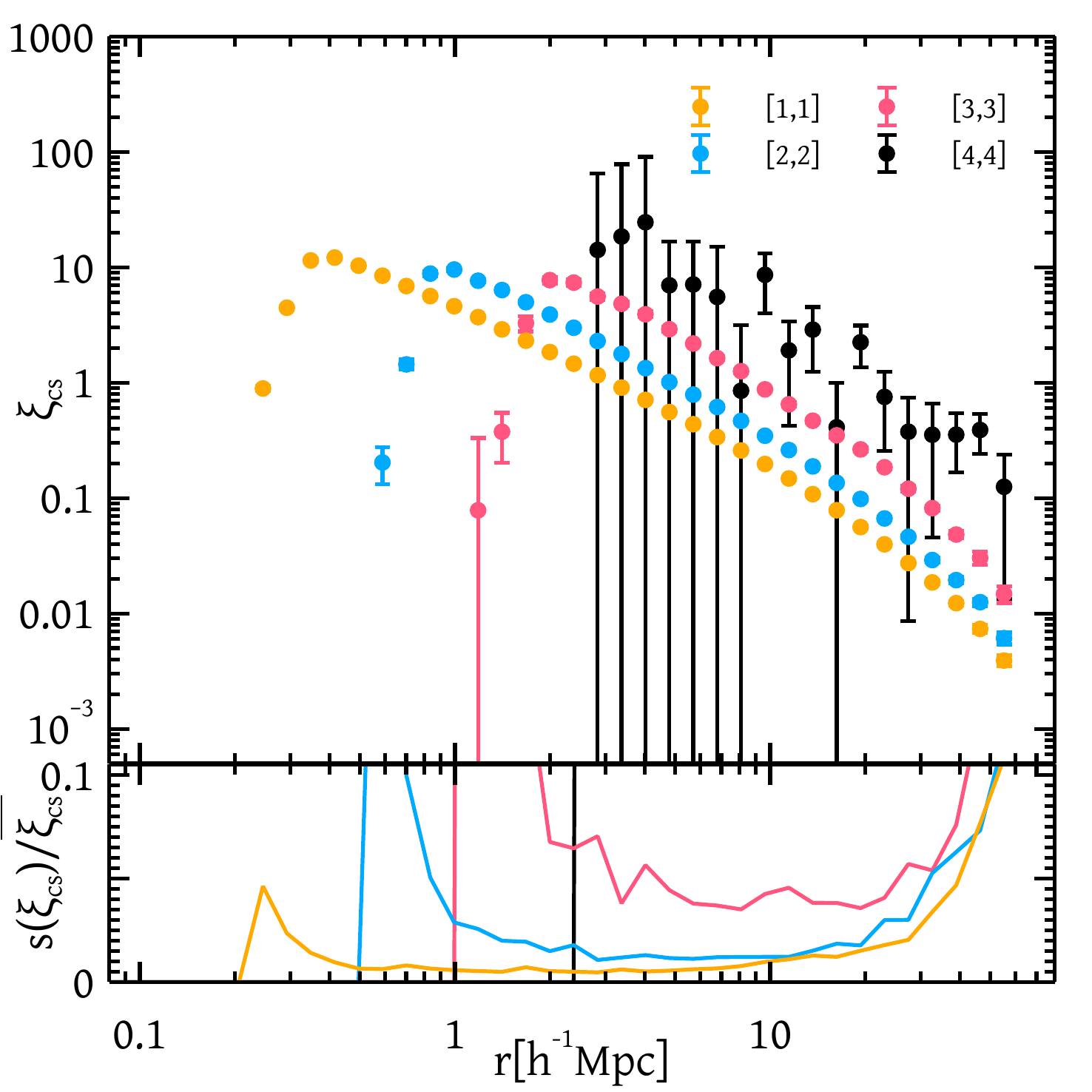} 
\caption{Similar to Figure~\ref{fig:rho_from_sim}, showing the central-satellite term $\xi_{\rm cs}$ of the $2$-h correlation function where both halos are in the same halo mass bin. The numbers inside square braces in the legend indicate the halo mass bins in the same order as shown in the legend of Figure~\ref{fig:rho_from_sim}.} \label{fig:xi_cs_diag_from_sim}
\end{figure}
\begin{figure}
\includegraphics[scale=0.5]{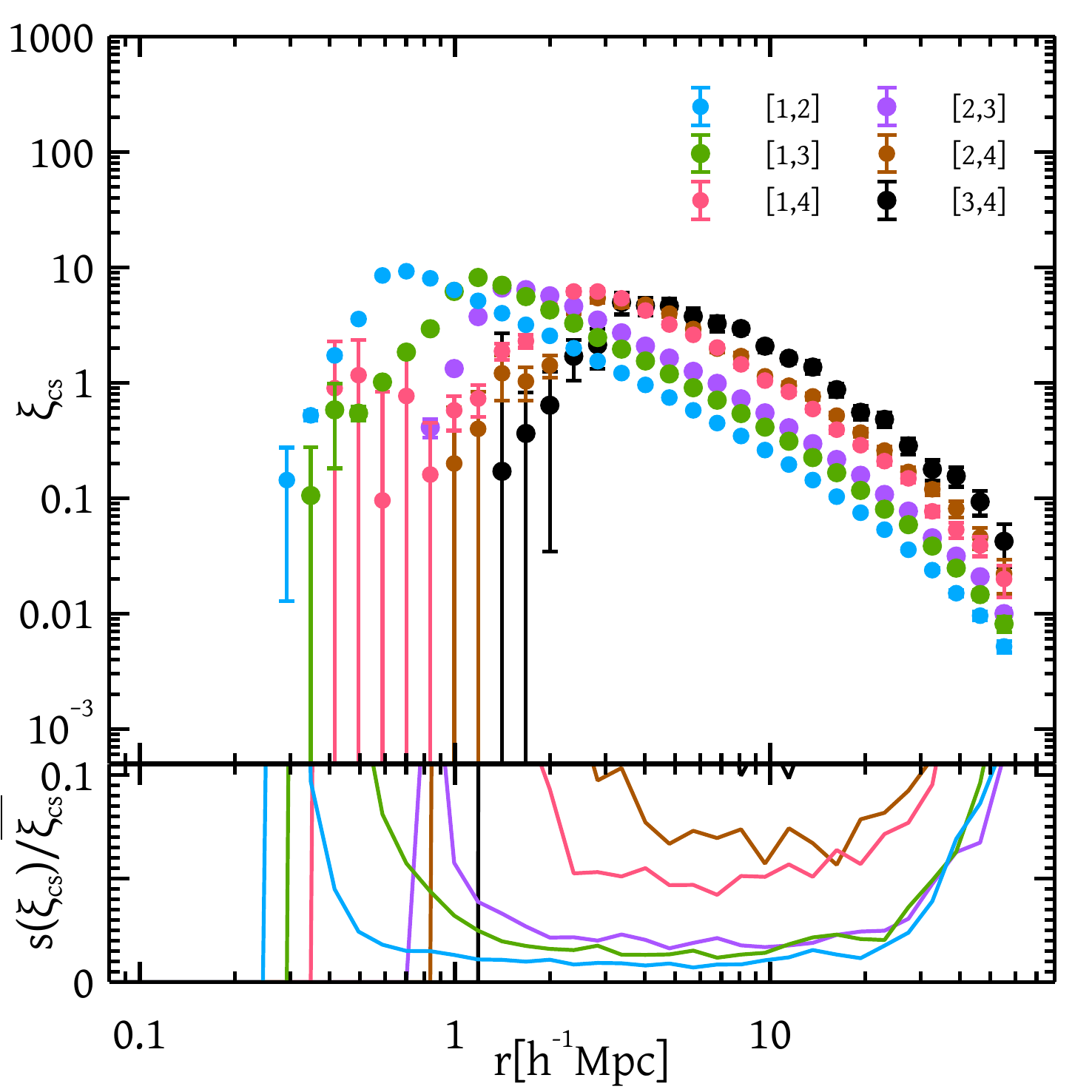} 
\caption{Similar to Figure~\ref{fig:xi_cs_diag_from_sim}, showing results for halos taken from two different halo mass bins, indicated by bin numbers in the legend using the same scheme as in Figure~\ref{fig:xi_cs_diag_from_sim}.} \label{fig:xi_cs_off_diag_from_sim}
\end{figure}

To reduce statistical noise, we took a weighted average of all the quantities weighted by the available number of particles in those measurements in each halo mass bin and in each realisation (for a detailed discussion of the weighting scheme, see Appendix~\ref{sec:simulation_weight}). The statistical variance on the mean value of our measurements is quite small as seen from the Figures~\ref{fig:hmf_from_sim} to \ref{fig:xi_cs_off_diag_from_sim}, which we discuss next.  

Figure~\ref{fig:hmf_from_sim} shows the weighted average of the halo mass function and its errors obtained from several realisations. As a comparison, in the lower panel of the figure, we have shown the ratio of the halo mass function to the fitting function of \citet{Tinker_2008_mass_function}; this shows agreement at the $\sim 5\%$ level over most of the mass range. We have also shown separate contributions of several boxes towards the halo mass function. 

Figure~\ref{fig:rho_from_sim} shows the measurements of halo density profile and the associated errors from different simulation boxes in different halo mass bins. We see that the errors on our measurements are $\lesssim3\%$ at all relevant scales. Figure~\ref{fig:lamda_from_sim} shows the convolution of the halo density profile with itself (which enters the $1$-halo term of the 2pcf) for different halo mass bins. We see that in this case also, the errors on the measurements are $\lesssim3\%$. As a sanity check on our measurements, we also show the NFW analytical forms of these quantities for halos of the same mass, using the median concentration-mass relation as measured from the simulation.\footnote{The \textsc{rockstar} code output contains information on the NFW scale radius which can be converted into a concentration for each halo, and we take the median concentration in each halo mass bin for use in the analytical curves.} We see good agreement at scales substantially larger than the force resolution of the respective simulation box. We emphasise that we only use the numerically measured and tabulated results for the halo profile in this work, not the NFW form.

Figure~\ref{fig:xi_cs_diag_from_sim} shows the central-satellite term $\xi_{cs}$ of the $2$-halo correlation function coming from halos within a single halo mass bin whereas Figure~\ref{fig:xi_cs_off_diag_from_sim} shows the same quantity measured from two different halo mass bins. We see from these plots that the errors on the measured quantities are also small, being $\lesssim5\%$ over nearly the entire range of masses and separations and rising to $\sim10\%$ or larger at the highest masses and largest separations. Similar trends are found for the other components of $2$-halo correlation function e.g. $\xi_{cc}$ and $\xi_{ss}$. For the sake of brevity, we do not show those plots here.

\subsection{Observational data set}
\label{sec:obs_data}
As observational data, we have used the projected 2pcf measurements of galaxies in the Sloan Digital Sky Survey\footnote{\href{http://www.sdss.org}{http://www.sdss.org}} \citep[SDSS, ][]{York_et_al_2000}  Data Release 7 \citep[DR7,][]{Abajajian_et_al_2009} as provided by \citet{Zehavi_et_al_2011}. These measurements were performed on a galaxy sample limited to an $r$-band Petrosian magnitude cut $r < 17.7$. This galaxy sample has a redshift range $0.02 < z < 0.25$ covering an area of $7700 \rm deg^2$ on the sky. The measurements are available for galaxies selected by absolute magnitude bins and thresholds, as well as for galaxies selected by Petrosian $g-r$ colour in absolute magnitude bins, split into red and blue populations in each magnitude bin using the cut
\begin{align}
(g-r)_{\rm cut} = 0.21 - 0.03 M_r \,\, , \label{eq:gr_cut}
\end{align}    
where $M_r$ is the Petrosian magnitude in the $r$-band, $K$-corrected and evolution corrected to redshift $z=0.1$. 

\begin{figure}
\includegraphics[scale=0.5]{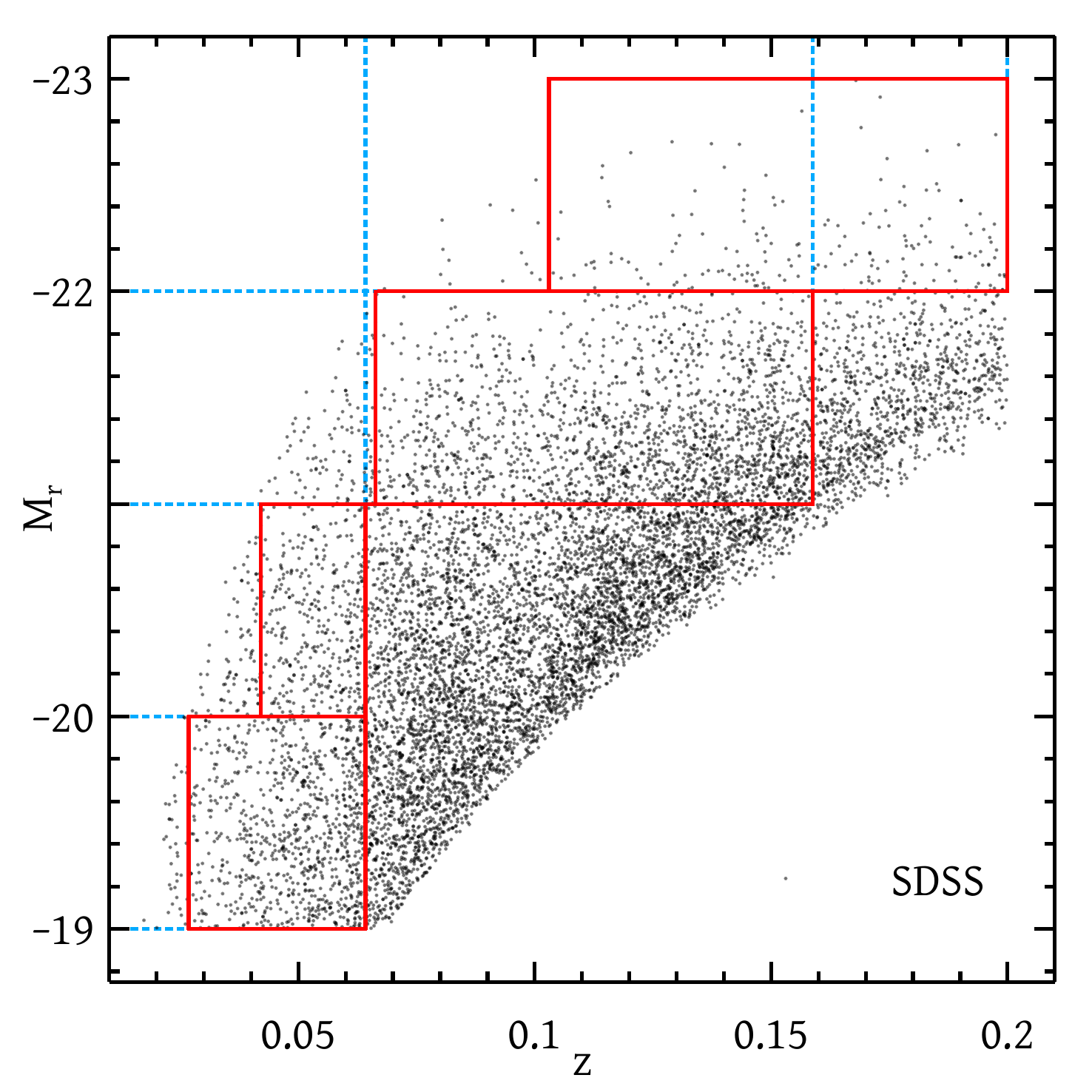} 
\caption{SDSS sample in the magnitude range $-23<M_r<-19$ and $14.5<m_r<17.7$, randomly resampled by a factor of 50. The red solid lines indicate the boundaries of volume-limited samples in different magnitude bins whereas the dotted blue lines indicate the boundaries for different magnitude thresholds. Noted that, for the magnitude bin $-21 < M_r < -20$, we have used a smaller upper redshift threshold than allowed by the data, in order to avoid the effect of Sloan Great Wall \citep[see Table 1 of][]{Zehavi_et_al_2011}. } 
\label{fig:SDSS_sample_threshold_bin} 
\end{figure}

Figure~\ref{fig:SDSS_sample_threshold_bin} shows the underlying galaxy luminosity data as a function of redshift, along with the various thresholds and bins of absolute magnitude for which 2pcf measurements are available. The tabulated measurements and associated covariance matrices have been kindly made public by I. Zehavi.\footnote{\href{http://astroweb.cwru.edu/izehavi/dr7\_covar/}{http://astroweb.cwru.edu/izehavi/dr7\_covar/}} 
Below we will describe a modification of these covariance matrices to account for the finite volume of the simulations used to build our theoretical model.

\subsection{Theoretical errors}
\label{sec:sim_errors}
To be able to model these correlation measurements accurately, the error from our simulation-based model should ideally be substantially smaller than the error associated with the observational measurements of the projected correlation function. To know the error coming from our model we need to know the HOD, which is not known \emph{a priori}. To break this circularity, we take the following approach. We first assume that the simulation errors are negligible compared to the data errors. We then determine the best-fit HOD using the global Markov Chain Monte Carlo (MCMC) procedure described in section~\ref{sec:global_HOD} below. For running the Monte Carlo chains, we used the package EMCEE\footnote{\href{http://dfm.io/emcee/current/}{http://dfm.io/emcee/current/}} \citep{Emcee_Mackey_et_al_2013}. Using this HOD, we can estimate the theoretical errors on the projected 2pcf which can be directly compared with the data errors. If these theoretical errors are smaller than the data errors, then our assumption of neglecting them is self-consistent.

\begin{figure}
\includegraphics[scale=0.5]{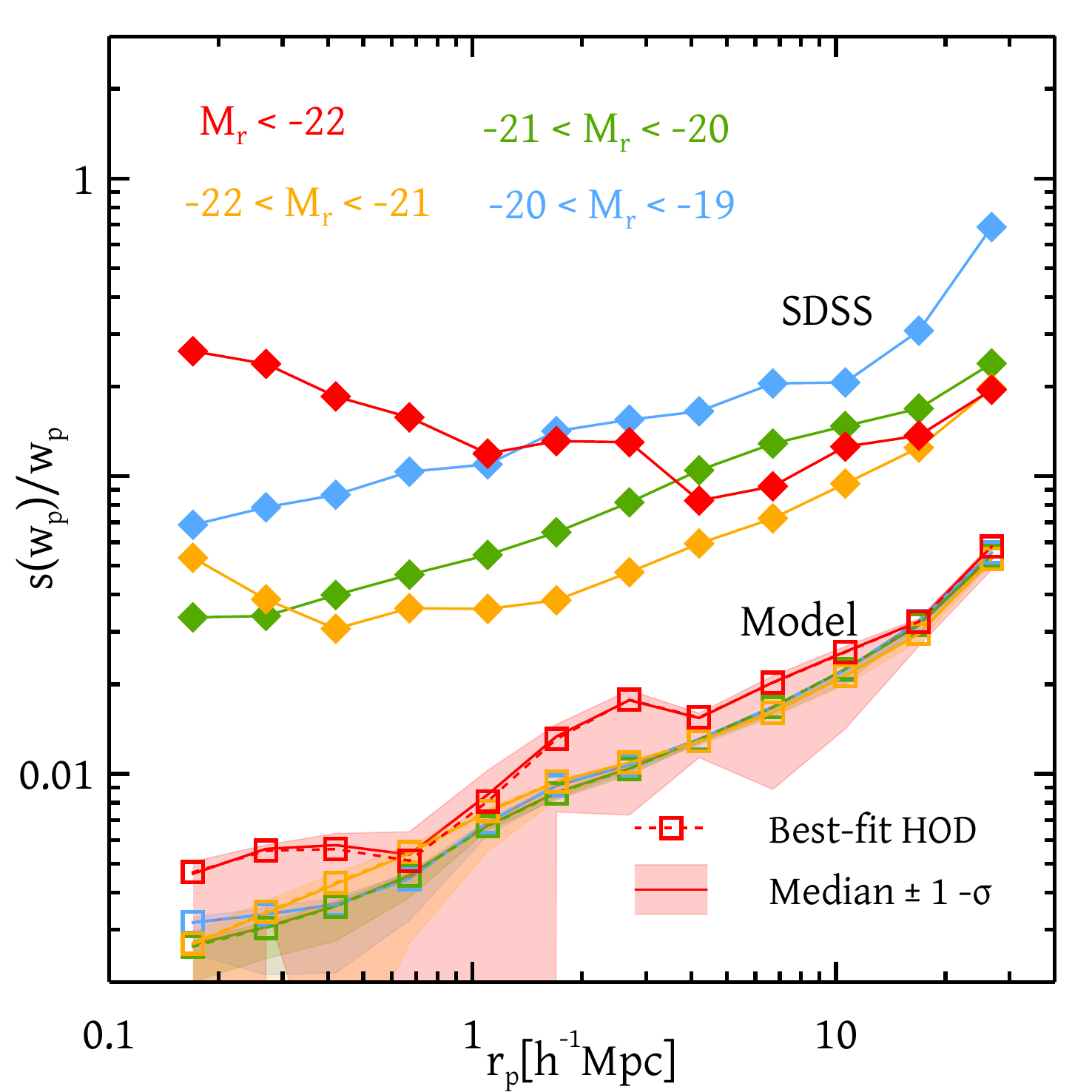} 
\caption{Relative errors on the projected correlation function from the observational dataset (filled diamond markers joined with solid lines) compared with the relative statistical errors from our simulation-based model using the best-fit HOD (empty square markers joined with dotted line) and the standard deviation of that relative error computed using 100 randomly sampled HOD-s from our MCMC chains (shown as error-bands). Different colours correspond to different magnitude thresholds or magnitude bins. We see that the model errors are always substantially smaller than the errors from the data. See main text for a detailed discussion of how the model errors are computed.} \label{fig:err_from_sim_all_bin_together}
\end{figure}

The detailed procedure of calculating the errors from our model can be found in Appendix~\ref{sec:simulation_weight}. While computing the errors, we assume that the weighted halo mass function and the average number of galaxies in a unit comoving volume do not have any error associated with them. For the $1$-halo term of the correlation function, in the halo mass range $10.6 < \log[m/\Mh] < 11.3$, we have the measurements from only one realisation of the $150 \Mpch$ box. So we assume a Poisson error over the measurements in that halo mass range. In the halo mass range $11.3 < \log[m/\Mh] < 14.0$, we take the $1$-halo measurements from $10$ realisations of the $300 \Mpch$ box and in the halo mass range $14.0 < \log [m/\Mh] < 15.5$, we take the $1$-halo measurements from $3$ realisations of the $600 \Mpch$ box. Computing the errors in the $1$-halo term (equation~\ref{eq:1h}) from several realisations of the boxes in these ranges of halo mass is straightforward. Then we add the errors coming from different ranges of halo mass in quadrature to compute the errors of the $1$-halo term over the full halo mass range. 

We take the $2$-halo measurements over the full halo mass range only from the $10$ realisations of the $300 \Mpch$ box and then compute the error using our best-fit HOD and equation~\eqref{eq:2h}. Now we add the errors of the $1$-halo and $2$-halo terms to compute the error of the full projected correlation function. In this way, we neglect any correlation between the $1$-halo and $2$-halo terms of the correlation function, which is justified because the effect of assembly bias is expected to be small.

We see in Figure~\ref{fig:err_from_sim_all_bin_together} that the errors from our model are indeed substantially smaller than the data errors over the entire relevant range of projected separations, thus justifying our choice of ignoring these theoretical errors in the HOD calibrations below.  Along with the relative error computed using the best-fit HOD, we have also shown $\pm 1$-$\sigma$ error band on that computed relative error by randomly selecting $100$ HOD-s from our MCMC chains. Since the errors coming from the measurements from simulations are very small compared to the errors coming from data, we neglect this error while constraining the parameters.

\section{Global analysis of projected clustering}
\label{sec:global_HOD}
In this section, we describe our global analysis of projected SDSS clustering and discuss the resulting constraints on the HODs and satellite red fraction.

\subsection{HOD Model} 
The halo occupation distribution (HOD) is a statistical model to populate halos with galaxies as a function of host halo mass. To compute the galaxy correlation function accurately in the halo model framework, one needs to split the galaxy population in centrals and satellites \citep{Zehavi_et_al_2005}. 

Following \citet{Zehavi_et_al_2011,Guo_et_al_2015}, we consider a five parameter based HOD approach where the two main quantities to model are $\fcen$ and $\Ns$. The quantity $\fcen(>L|m)$ denotes the fraction of $m$-halos (halos having mass in the range $m$ to $m + \der m$) which contain a central galaxy with luminosity greater than $L$. The other quantity $\Ns(>L|m)$ denotes the average number of satellites with luminosity greater than $L$ in an $m$-halo having a central galaxy with luminosity greater than $L$. We model these two quantities in the following way,
\begin{align}
\fcen (>L|m) &= \frac{1}{2} \left[1 + \erf{\frac{\log m - \log M_{\rm min}}{\sigma_{\rm \log M}}} \right] \label{eq:fcen} \\
\Ns(>L|m) &= \left( \frac{m - M_0}{M_1^{\prime}} \right)^{\alpha} \label{eq:Ns} \,\, .
\end{align}
From the above equations, it is clear that in our model,  we have a total of five free parameters: the cut-off mass cale $M_{\rm min}$,  width of the central galaxy mean function  $\sigma_{\log M}$ and cut-off mass scale $M_0$, normalization $M_1^{\prime}$ and high mass slope $\alpha$ of the mean occupation function of the satellite galaxies. 

Once we have modelled the population of galaxies in a halo, we can convolve this with halo statistics and obtain the statistics of galaxies. The average number density of galaxies can be computed from the halo number density $n(m_i)$ as,
\begin{align}
\bar{n}_g = \Sigma_{i = 1}^{N} \left[ \fcen(m_i) + \Nscript (m_i) \right] n(m_i) \der \log m_i \,\, . 
\end{align}

Similarly we can also compute the $1$-halo and $2$-halo term of the correlation function as follows, 
\begin{align}
\xi_{gg}^{1h} (r) &= \sum_{i = 1}^{N} \der \log m_i n(m_i) \frac{\fcen (m_i)}{\bar{n}_g^2} \left[ 2 \Ns(m_i) \frac{\rho(r|m_i)}{m_i} \right. \notag \\
& \left. + \Ns^2(m_i)  \frac{\lambda(r|m_i)}{m_i^2} \right] \,\, .
\end{align}  
and
\begin{align}
\xi_{gg}^{2h}(r) &= \sum_{i = 1}^{N} \sum_{j = 1}^N \der \log m_i \der \log m_j \frac{n(m_i) n(m_j)}{\bar{n}_g^2} \notag \\
& \times \left[ \fcen(m_i)\fcen(m_j) \xi_{hh}^{cc}(r|m_i, m_j) + 2 \fcen(m_i)  \Nscript (m_j) \right. \notag \\
& \left. \times \xi_{hh}^{cs} (r|m_i, m_j) + \Nscript(m_i) \Nscript(m_j) \xi_{hh}^{ss}(r|m_i, m_j)  \right] \,\, , \label{eq:2h}
\end{align}
To get a detailed discussion about this formalism please see appendix~\ref{sec:2PCF_calculation_formulae}.
\begin{figure}
\includegraphics[scale=0.5]{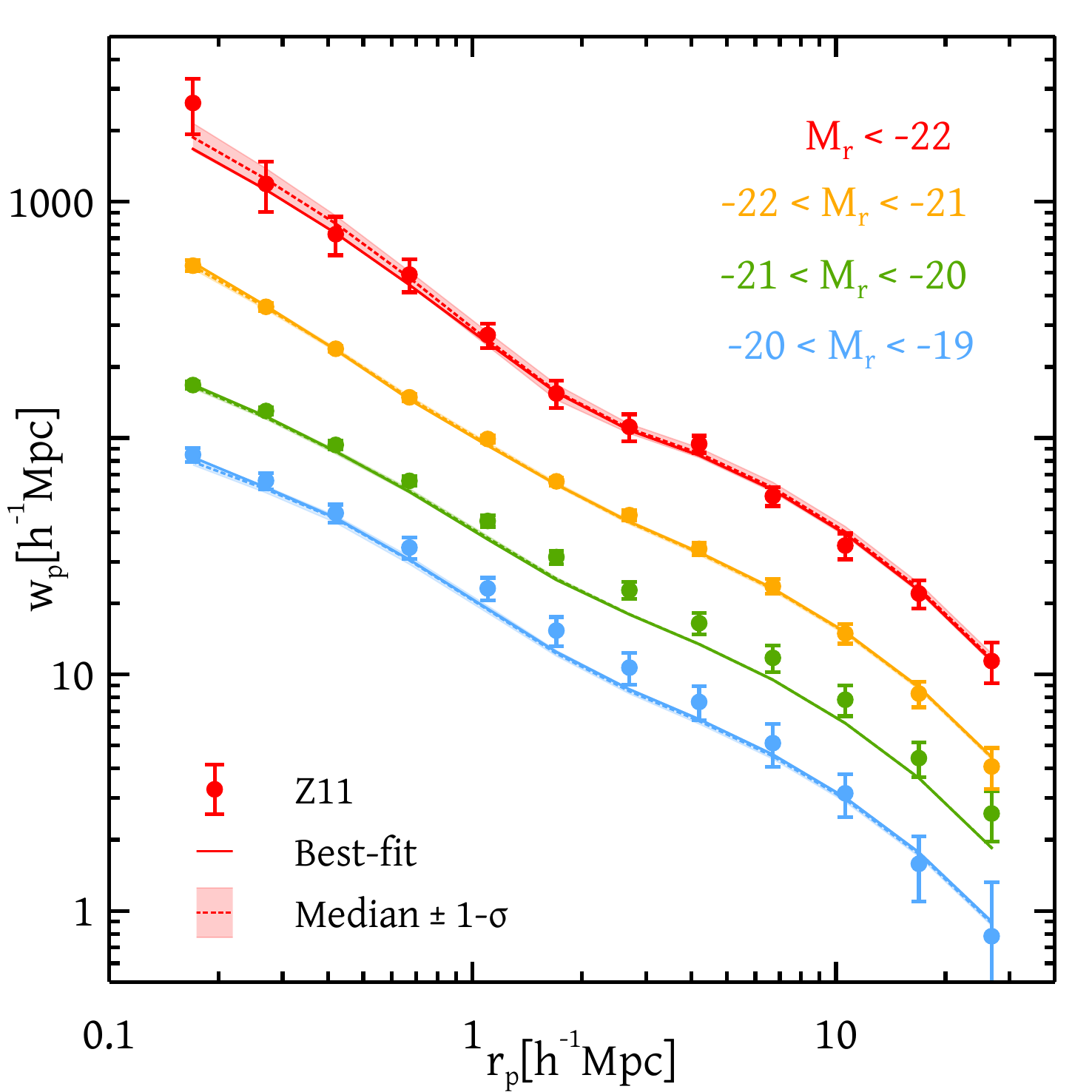} 
\caption{Projected correlation function of luminosity-selected galaxy samples. Solid circles with error bars show the measured data and solid curves show the best-fit correlation functions computed from our model. The median and $\pm 1$-$\sigma$ regions for the model are shown with dashed lines and an error band. Starting from the top, the last two magnitude bins are artificially separated by $0.25$dex. See main text for a discussion.}
\label{fig:corr_all_color_correlation_function}
\end{figure}

While doing the global analysis of HOD-s of all magnitude bins together as discussed in the next subsection, we took the following approach to ensure that the HODs of dimmer magnitude thresholds gives a higher (or equal) number density than brighter thresholds. While running the MCMC, if for some choices of parameters, the HOD-s for a given magnitude bin which is the difference of HOD-s of two adjacent magnitude thresholds become negative in a certain halo mass range, we forcefully assign the binned HOD-s to be zero in that halo mass range. Since we are also using the data of number density of galaxies in different magnitude bins in our analysis, which has already the trend of having higher number density in case of dimmer magnitude bins, the HOD-s in our model already get constrained to follow that trend. 

Additionally, while modelling colour-dependent clustering, we also include the red fraction of satellite galaxies $p_{\rm rs}$ as a free parameter. A detailed discussion on calculating colour-independent and colour-dependent 2pcfs can be found in Appendix~\ref{sec:2PCF_calculation_formulae}. 

\subsection{Likelihood calculation}
Since the galaxy samples of different magnitude thresholds are highly correlated as seen from Figure~\ref{fig:SDSS_sample_threshold_bin}, it is cleaner to use the measurements from different magnitude bins for the HOD calibration. However, the HOD parametrisation for magnitude thresholds is more well-established than the one for magnitude bins. The HODs in different luminosity bins ($L_{12}$) can be computed from HODs of adjacent luminosity thresholds ($L_1$ and $L_2$) in a straight-forward way as shown below,
\begin{align}
\fcen(L_{12}|m) &= \fcen(>L_1|m) - \fcen(>L_2|m) \,\, \\
\Nscript(L_{12}|m) &= \Nscript(>L_1|m) - \Nscript(>L_2|m) \,\, , \\
\Ns(L_{12}|m) &= \Nscript(L_{12}|m)/\fcen(>L_1|m) \,\, . 
\end{align}
The correlation function as a function of magnitude bin can be computed from those derived binned HODs as discussed in detail in the Appendix~\ref{sec:HOD_fomulae_in_bin}. 

The HOD of an individual magnitude threshold will contribute to the measurements of two adjacent magnitude bins. To constrain the HODs of magnitude thresholds correctly, therefore, we can construct a global likelihood using the \emph{uncorrelated measurements} from a range of contiguous magnitude bins, which then allows us to simultaneously constrain the parameters corresponding to all thresholds, as we describe next. 

\begin{figure}
\includegraphics[scale=0.5]{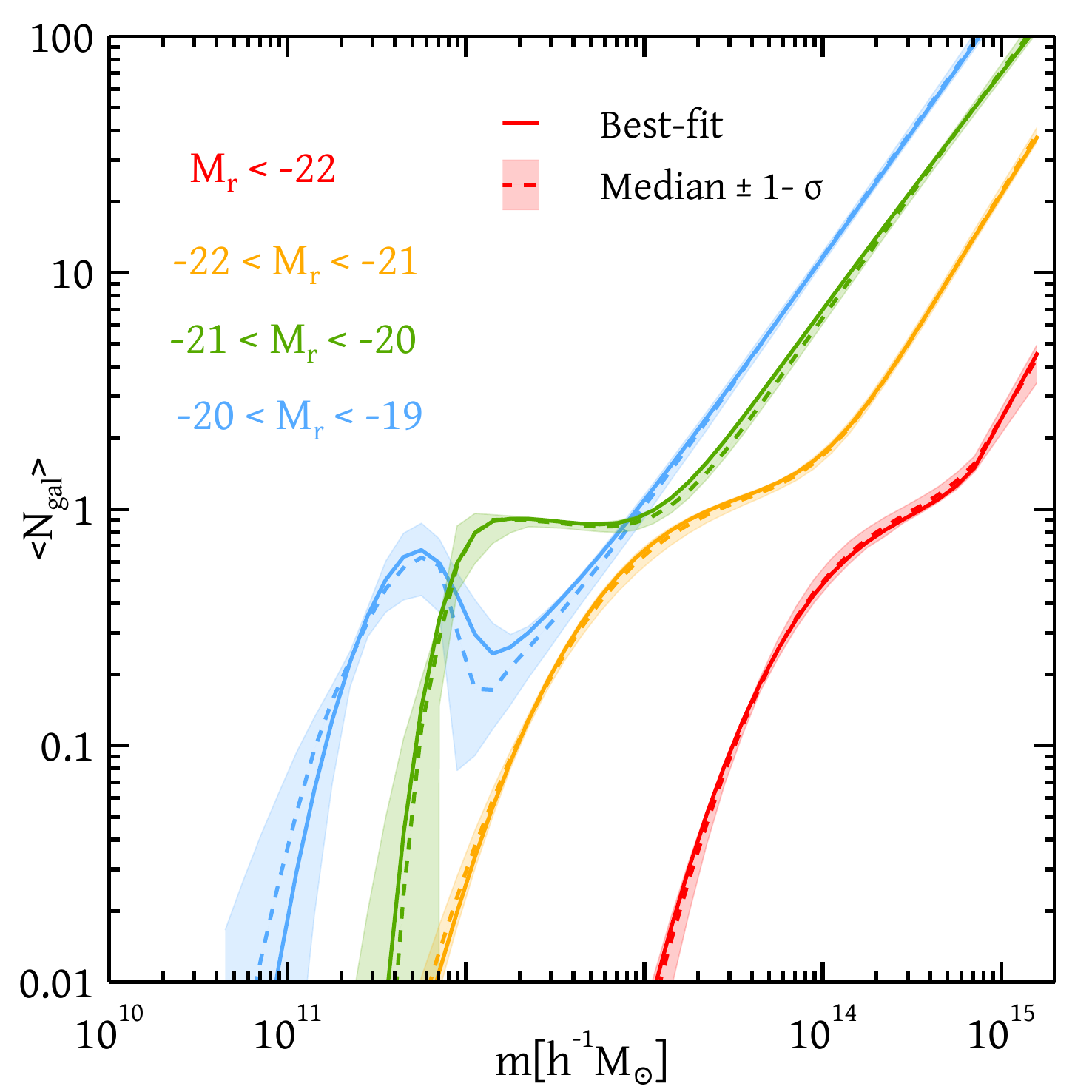} 
\caption{Best-fit HODs (solid curves) as a function of halo mass for different magnitude thresholds and bins.  Dashed curves with error bands show the median and $\pm 1$-$\sigma$ error on the HOD.}
\label{fig:HOD_all_color}
\end{figure}

For galaxies in a luminosity bin $l$, we assume a Gaussian likelihood expressed as
\begin{align}
\ln \mathcal{L}_l({\theta}) &= -\frac{1}{2} \left(\hat{\bf{w}}_p - {\bf{w}}_p(\theta)\right)^T \tilde{C}^{-1} \left(\hat{\bf{w}}_p - {\bf{w}}_p(\theta)\right) \notag\\
&\ph{\left({\bf{w_p}}^* - {\bf{w_p}}(\theta)\right)^T}
- \frac{1}{2}\frac{\left(\hat n_g - n_g(\theta)\right))^2} {\sigma_g^2} \,, \label{eq:lnlikelihood-lum}
\end{align}
where ${\bf{w_p}}$ is the vector of measurements of the projected correlation function of galaxies in this luminosity bin, $\tilde{C}^{-1}$ is the modified inverse of the corresponding covariance matrix (see below) and $n_g$ is the average comoving number density of galaxies in this bin with the associated error $\sigma_g$. Quantities with carets ($\,\hat{}\,$) are measurements from observations and the ones without denote the model prediction which depends on the parameter set denoted by $\theta$. \emph{We combine measurements from the available luminosity bins by summing over the corresponding log-likelihoods.} Essentially, our analysis uses a Bayesian technique where the posterior probability distribution function is a product of the likelihood function and prior distribution of parameters. In each luminosity bin, we have five free parameters $\log M_{\rm min}$, $\sigma_{\log M}$, $\log M_0$, $\log M_1^{\prime}$ and $\alpha$ . For each of them we choose a flat uniform prior. In each bin, the prior range for $\log M_{\rm min}$, $\log M_0$ and $\log M_1^{\prime}$ was chosen to be [10.65, 15.5], the prior for $\sigma_{\log M}$ was taken as $[0.001, 5.0]$ and for $\alpha$ it was taken to be $[0.001, 5.0]$.

\citet{Zehavi_et_al_2011} also provide clustering measurements for red and blue galaxies separately in different luminosity bins. When including this information, \emph{we use the all-colour constraints using \eqn{eq:lnlikelihood-lum} as a prior}, combined with the following likelihood for the $l^{\rm th}$ luminosity bin
\begin{align}
\ln \mathcal{L}_l({\theta}) &=  -\frac{1}{2} \left(\hat{\bf{w}}_p^{\rm (r)} - {\bf{w}}_p^{\rm (r)}(\theta)\right)^T \tilde{C^{\rm (r)}}^{-1} \left(\hat{\bf{w}}_p^{\rm (r)} - {\bf{w}}_p^{\rm (r)}(\theta)\right) \notag\\
& -\frac{1}{2} \left(\hat{\bf{w}}_p^{\rm (b)} - {\bf{w}}_p^{\rm (b)}(\theta)\right)^T \tilde{C^{\rm (b)}}^{-1} \left(\hat{\bf{w}}_p^{\rm (b)} - {\bf{w}}_p^{\rm (b)}(\theta)\right)\,, \label{eq:lnlikelihood-col}
\end{align}
where the superscripts (r) and (b) refer to red and blue galaxies, respectively. As before, we combine results from different luminosity bins by summing over the respective log-likelihoods. While using this extra information of colour dependent clustering, we need to incorporate one extra parameter $p_{\rm rs}$ (the satellite red fraction) in each luminosity bin to model the colour-dependent 2PCF correctly. For each $p_{\rm rs}$ parameter we choose a flat uniform prior in the range $[0.01, 0.99]$.

\begin{figure}
\includegraphics[scale=0.5]{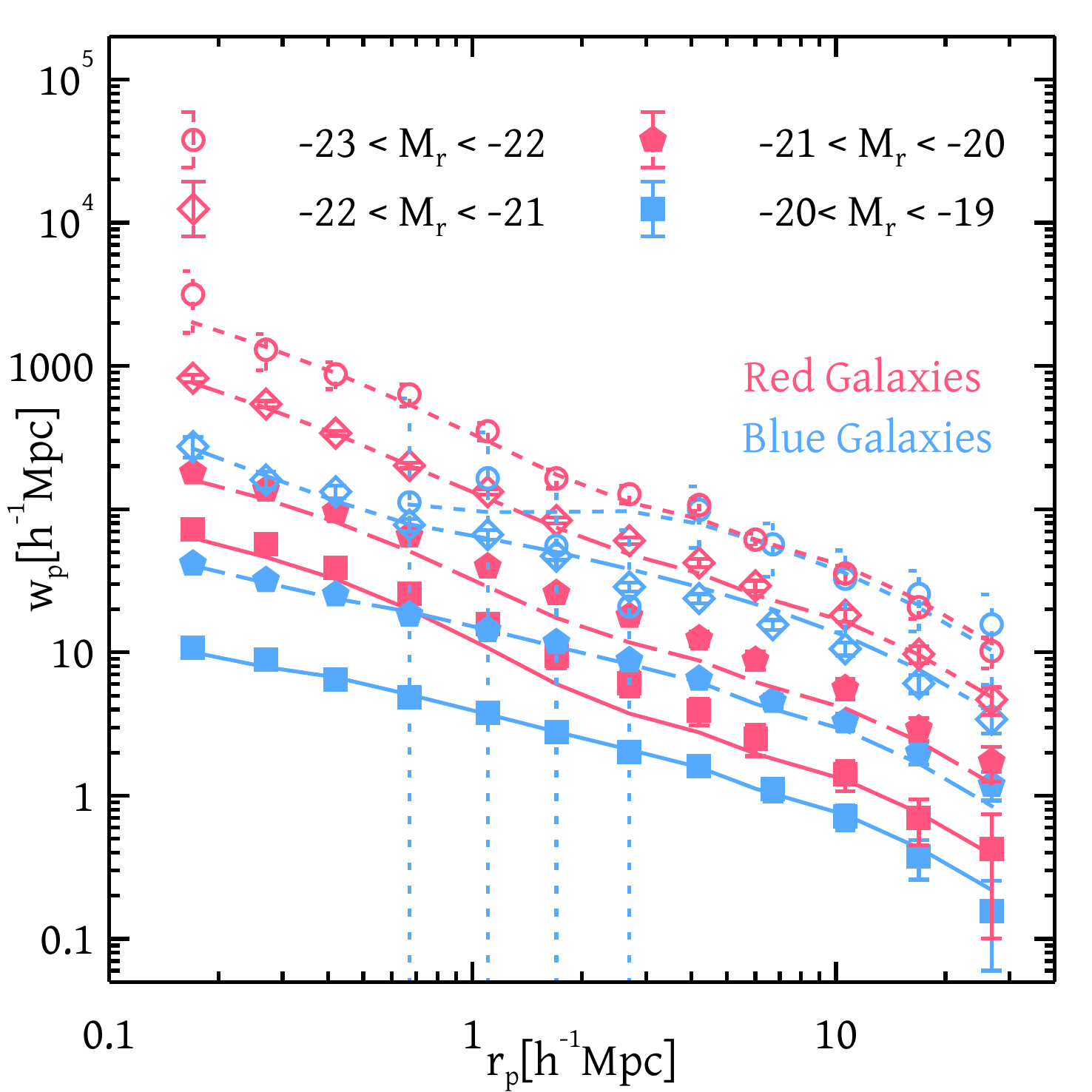} 
\caption{Projected correlation function of red and blue galaxies separately for different magnitude bins. Points with error bars show the measured data while the curves show our best-fit model. Starting from the top, the data points and the curves for the last two magnitude bins are artificially separated by $0.5$dex for clarity.} \label{fig:corr_red_blue}
\end{figure}

\subsection{Data set and covariance matrices}
Following \citet{Guo_et_al_2015},  we compute all of $\tilde{C}^{-1}$, $\tilde{C^{\rm (r)}}^{-1}$ and $\tilde{C^{\rm (b)}}^{-1}$ in two steps. At first, to account for the model uncertainties arising due to finite volume effect of the simulations, we multiply the covariance matrix measured from the data with the factor $\left(1 + V_{\rm data}/V_{\rm sim}\right)$.  For every magnitude bin, we have a $z_{\rm min}$ below which we do not have any galaxy brighter than the bright end of that magnitude bin and a $z_{\rm max}$ above which we do not have any galaxy fainter than the faint end of the magnitude bin. From these two redshift bounds, one can easily compute the maximum and minimum comoving distance ($\chi$) of the galaxies from us. So $V_{\rm data} = {\rm sky \,\, area} \times (\chi_{\rm max}^3 - \chi_{\rm min}^3)/3$. 

Since our correlation measurements for all halo mass range come mainly from the $300\Mpch$ box, to be conservative, we have taken $V_{\rm sim} = 10 \times 300^3 h^{-3} {\rm Mpc}^3$. Now to get an unbiased estimate of the inverse of this new covariance matrix we multiply its inverse with the factor of $(1-D)$, where $D \equiv (n_{\rm d} + 1)/(n_{\rm jack} -1)$ \citep{Hartlap_et_al_2007, Percival_et_al_2014}. In this expression $n_{\rm d}$ denotes the number of data points and $n_{\rm jack}$ denotes the number of jackknife samplings used to compute the error covariance matrix. Following \citet{Zehavi_et_al_2011}, we have assumed a $5 \%$ error in the measurement of $\bar{n}_g$.

It is worth mentioning here that while fitting colour-\emph{independent} clustering and abundance data, we fit the data for the brightest magnitude threshold, i.e. $M_r < -22$ and next three magnitude bins i.e. $[-22, -21], [-21, -20]$, and $[-20, -19]$. While fitting the colour-\emph{dependent} clustering data, on the other hand, we considered the red and blue correlation measurements from four magnitude bins $[-23, -22], [-22, -21], [-21, -20]$ and $[-20, -19]$. We also assumed our HOD for the $M_r < -22$ threshold to be equal to the HOD for the brightest magnitude bin. This is justified because there are few very bright galaxies with $M_r < -23$ compared to those with $-23 < M_r < -22$. 

\begin{figure}
\includegraphics[scale=0.5]{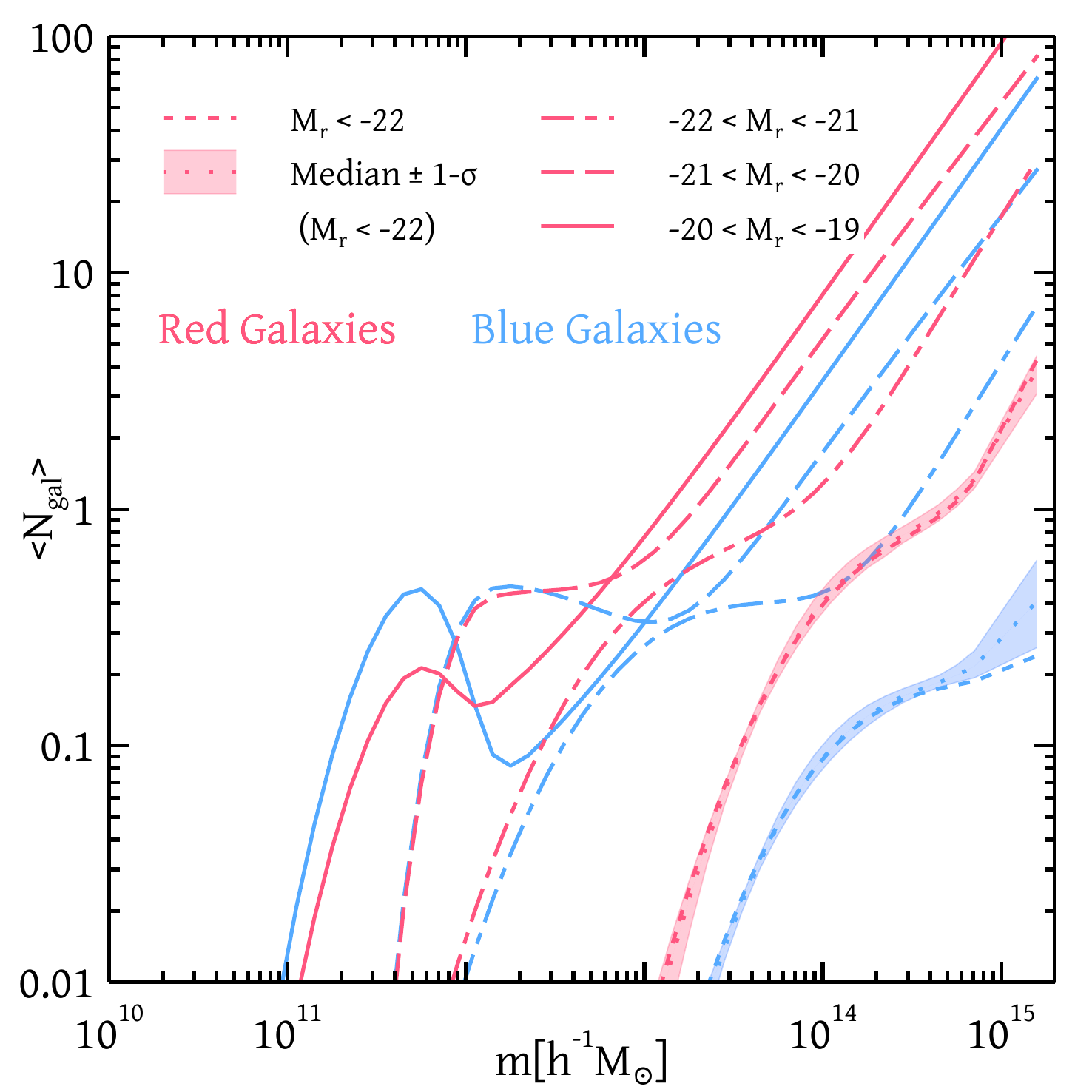} 
\caption{Best fit HODs of red and blue galaxies separately for different magnitude bins. For clarity, we have shown the median and $\pm1$-$\sigma$ error bands only for the brightest sample. The errors for the other luminosity bins are qualitatively similar. }
\label{fig:HOD_red_blue}
\end{figure}

We restrict our analysis to projected separations $r_{\rm p}\lesssim40\Mpch$, since our simulation box volumes do not allow us to reliably probe larger projected scales due to the absence of long-wavelength modes. Following \citet{Zehavi_et_al_2011}, we have assumed $ \pi_{\rm max} = 60 \Mpch$ in all the calculations in our model. 

\begin{figure*}
\includegraphics[width=0.3\textwidth]{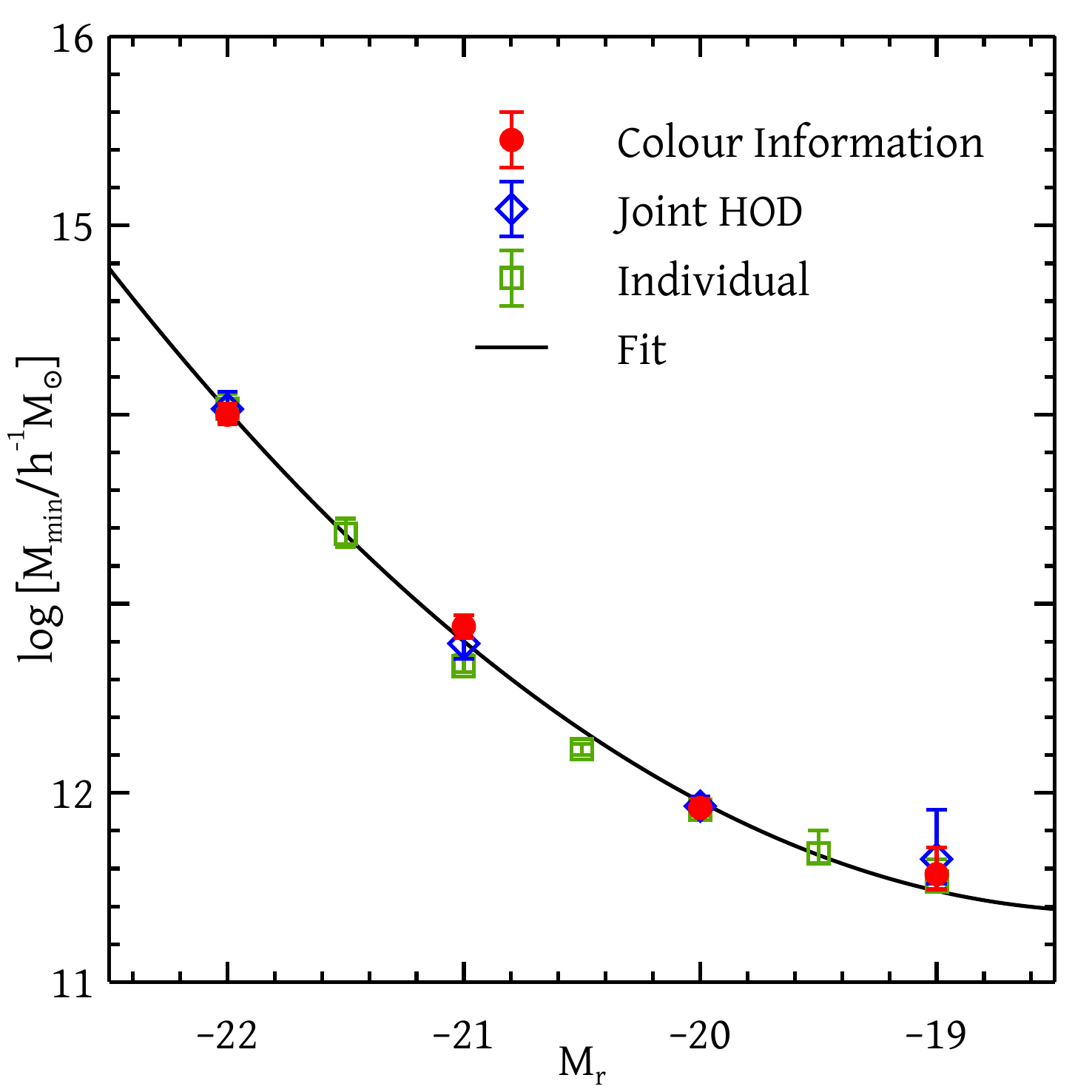} 
\includegraphics[width=0.3\textwidth]{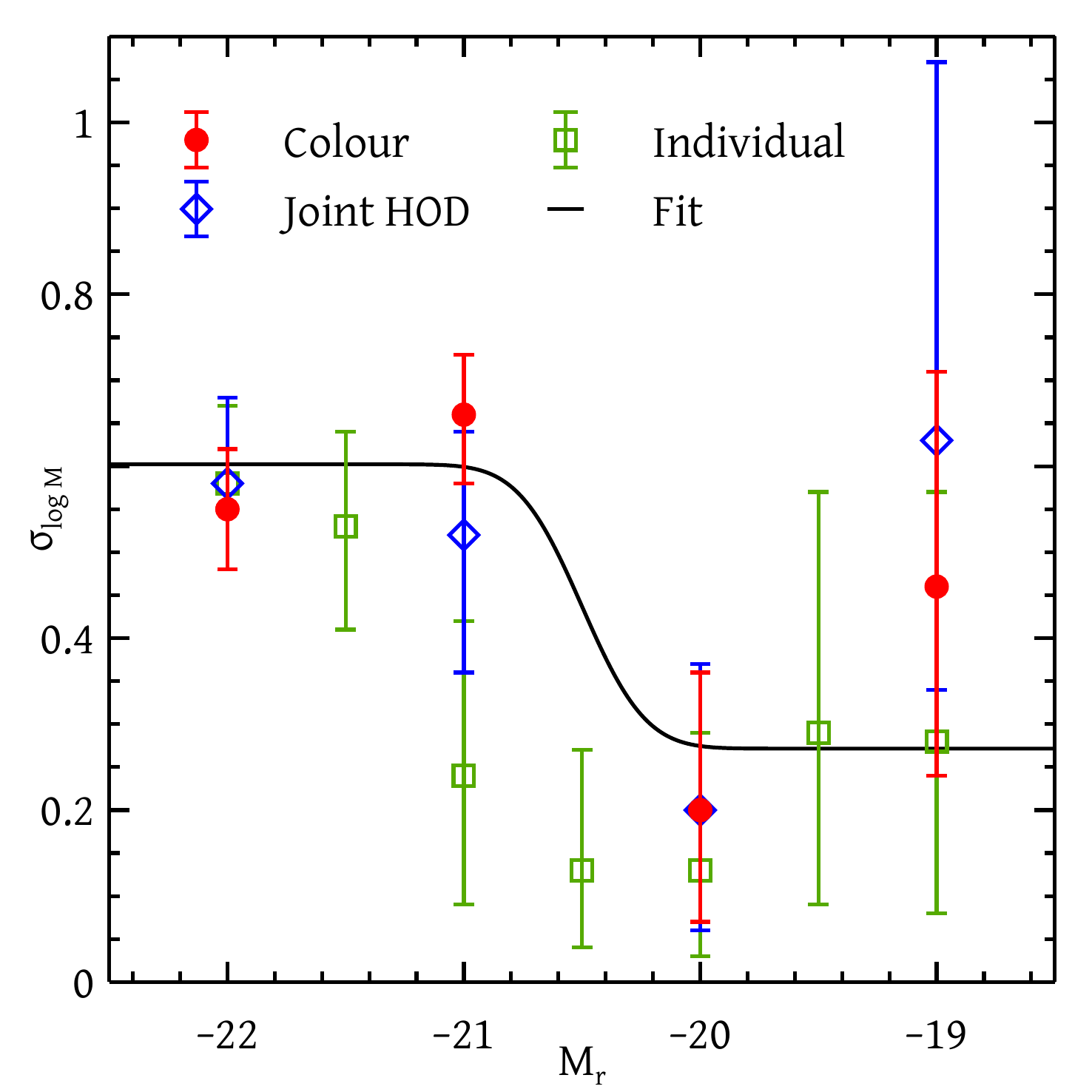} 
\includegraphics[width=0.3\textwidth]{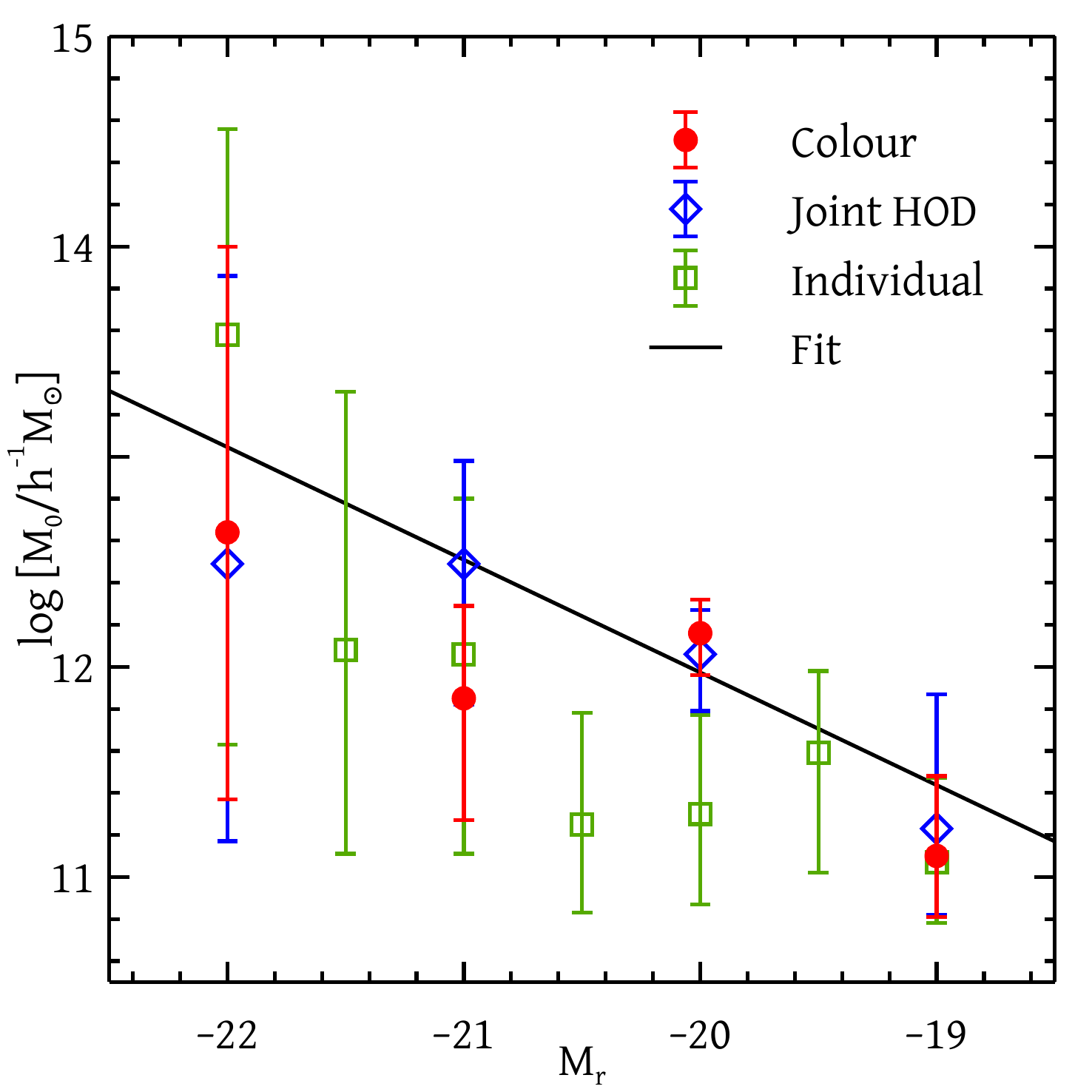} 
\caption{HOD parameters as a function of magnitude thresholds: $\log M_{\rm min}$ \emph{(left panel)}, $\sigma_{\log M}$ \emph{(middle panel)} and $\log M_{0}$ \emph{(right panel)}.
The solid red circles with error bars show the parameter values obtained after a global HOD analysis along with the information of colour-dependent clustering; this is our default set of constraints. The open blue diamonds with error bars show the same as obtained without colour information. The green open squares with error bars show constraints using 2pcf measurements of individual magnitude thresholds separately, as is usually done in the literature. The solid black lines show our best-fit estimates for the fitting functions described in Table~\ref{tab:HOD_fit} using the red circles.}
\label{fig:log_Mmin_fit}
\end{figure*}

\begin{figure*}
\includegraphics[width=0.3\textwidth]{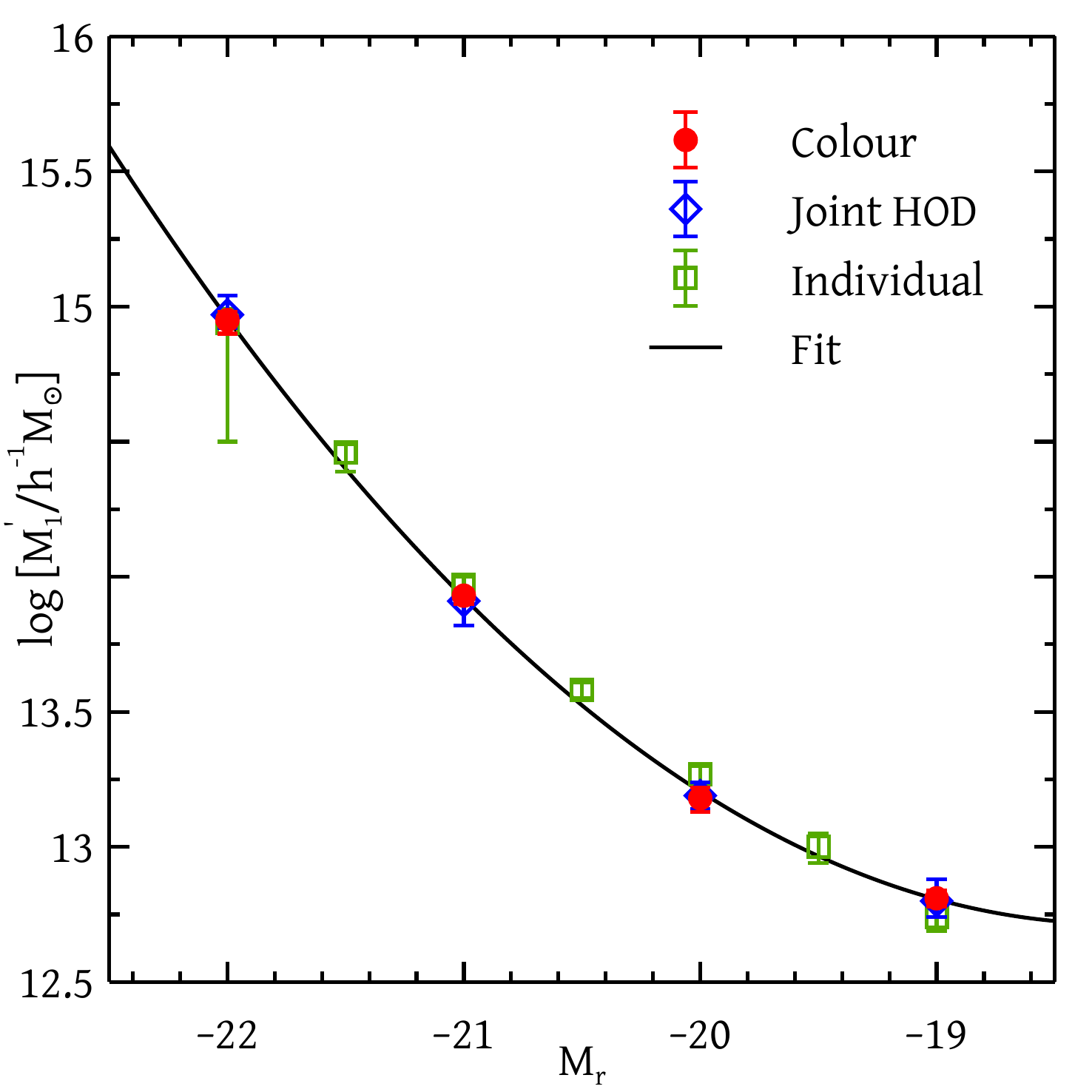} 
\includegraphics[width=0.3\textwidth]{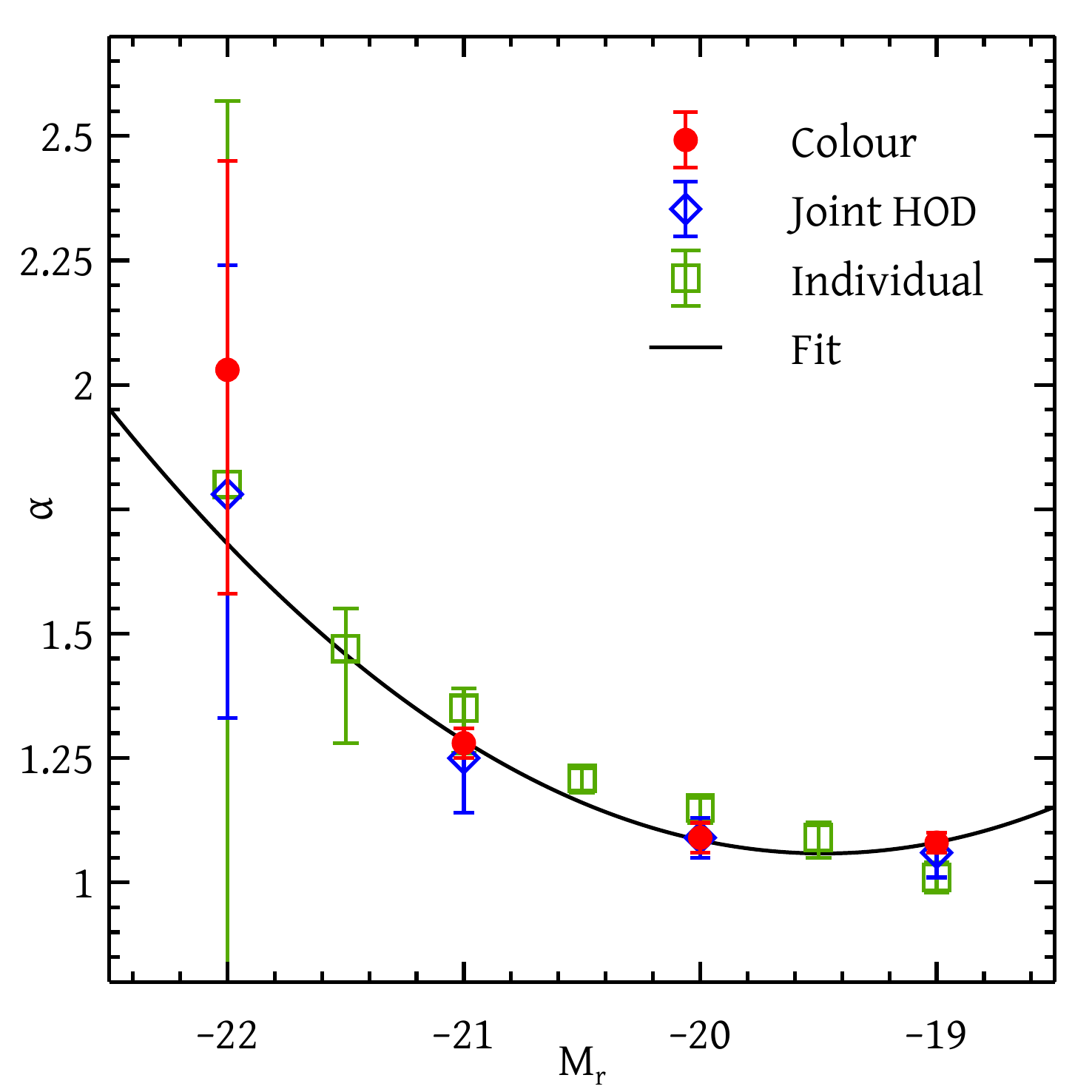} 
\includegraphics[width=0.3\textwidth]{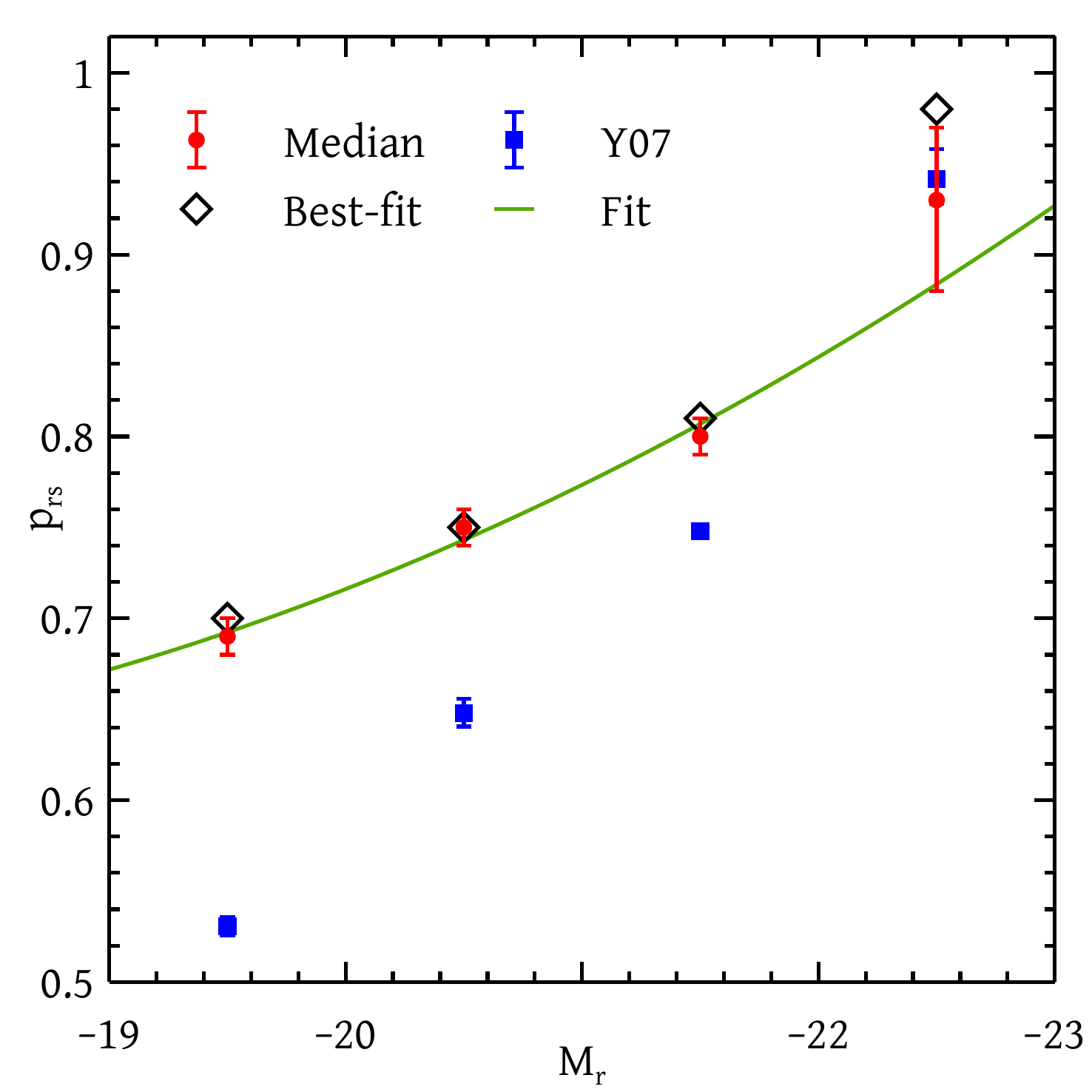} 
\caption{The \emph{left} and \emph{middle panels} of this Figure are the same as Figure~\ref{fig:log_Mmin_fit} but for the HOD parameters $\log M_1^{\prime}$ and $\alpha$, respectively.  The \emph{right panel} shows the luminosity dependence of the satellite red fraction $p_{\rm rs}$. The red points with error bars show the median value of satellite red fraction obtained from our calibrations of colour-dependent clustering analysis, while the black open diamonds show the best-fit results. The blue squares show the measurements from the group catalog of \citet[][Y07]{Yang_et_al_2007}. The green solid curve shows the best-fit function of the satellite red fraction as described in Table~\ref{tab:HOD_fit}.} \label{fig:alpha}
\end{figure*}
%
\subsection{Convergence of the chains}
In this section, we briefly discuss the convergence criterion for our MCMC chains. For the colour-independent HOD formalism, we ran a total of $1024$ walkers with each chain run for $100000$ steps. For each walker we throw away first $2000$ steps as a burn-in. Then we compute the correlation length of the remaining samples using EMCEE-s default \textit{autocorr} function. Then we join the chains and thin them with the proper correlation length to get independent samples of our posterior distribution. We also run the chains for $5000$ steps and $15000$ steps and the results were same. So we concluded that $10000$ steps with $1024$ walkers are sufficient enough for the walkers to converge. Traditional methods like Gelman-Rubin test etc. are not appropriate in case of EMCEE because all the walkers are correlated.

For the colour-dependent clustering analysis, we drew the priors from the posteriors of the previous colour-independent analysis. In this sampling, we ran $1024$ walkers with $5000$ steps each. We followed same procedure as discussed above for computing the correlation length and then thin the samples by the correlation length. We also did the same analysis with $3000$ steps for each walker and again for $10000$ steps per walker. The results were very similar. So we concluded then that $1024$ walkers with $5000$ steps each was good enough to achieve the convergence.
\section{Results}    
\label{sec:results}
\noindent
In this section, we present the results of our MCMC fitting exercise along with fitting functions for the HOD parameters and satellite red fraction as a function of luminosity and discuss a few potential applications.

\begin{table}
\centering
\begin{tabular}{c c}
$M_r$ bin & $p_{\rm rs}$ \\
\hline \hline 
$[-23, -22]$ & $0.98 (0.93^{+0.04}_{-0.05})$\\ 
$[-22, -21]$ & $0.81 (0.80^{+0.01}_{-0.01})$\\
$[-21, -20]$ & $0.75 (0.75^{+0.01}_{-0.01})$\\
$[-20, -19]$ & $0.70 (0.69^{+0.01}_{-0.01})$\\
\hline
\end{tabular}
\caption{Satellite red fraction $p_{\rm rs}$ for different magnitude bins obtained as described in the text. We quote the best-fit values of the parameters (outside parentheses) as well as the median and $\pm 1$-$\sigma$ values (inside parentheses).}
\label{tab:prs_joint_fit}
\end{table}

\begin{table*}
\centering
\begin{tabular}{c | c | c | c | c | c}
$M_{r}^{\rm max}$ & $\log M_{\rm min}$ &$\sigma_{\log M}$ & $\log M_0$ & $\log M_1^{\prime}$ & $\alpha$ \\ 
\hline \hline 
\rule{0pt}{3ex}  -22 & $14.02 (14.00^{+0.06}_{-0.05})$ & $0.58 (0.55^{+0.07}_{-0.07})$ & $12.55 (12.64^{+1.36}_{-1.27})$ & $14.96 (14.95^{+0.03}_{-0.05})$ & $2.35 (2.05^{+0.42}_{-0.45})$ \\
\hline 
\rule{0pt}{3ex} -21 & $12.85 (12.88^{+0.06}_{-0.06})$ & $0.62 (0.66^{+0.07}_{-0.08})$ & $11.36 (11.85^{+0.44}_{-0.58})$ & $13.93 (13.93^{+0.02}_{-0.03})$ & $1.27 (1.28^{+0.03}_{-0.03})$ \\
\hline
\rule{0pt}{3ex} -20 & $11.91 (11.92^{+0.05}_{-0.03})$ & $0.22 (0.20^{+0.16}_{-0.13})$ & $12.21 (12.16^{+0.16}_{-0.20})$ & $13.12 (13.18^{+0.04}_{-0.05})$ & $1.05 (1.09^{+0.03}_{-0.03}) $ \\
\hline
\rule{0pt}{3ex} -19 & $11.55 (11.57^{+0.14}_{-0.08})$ &  $0.38 (0.46^{+0.25}_{-0.22})$ &  $10.91 (11.10^{+0.38}_{-0.29})$ & $12.78 (12.81^{+0.03}_{-0.03})$ & $1.06 (1.08^{+0.02}_{-0.02})$ \\
 \hline
\end{tabular}
\caption{HOD parameters for different magnitude thresholds obtained using colour information (see main text for a discussion). The values outside parentheses show the best-fit results of our global analysis, while the median and $\pm 1$-$\sigma$ values are quoted inside parentheses. For this analysis, we found the total $\chi^2/dof = 142.05/117$. } \label{tab:HOD_global_with_colour}
\end{table*}

\subsection{Luminosity-dependent clustering}
We have initially performed our analysis for luminosity-dependent clustering without accounting for any colour information. Using the results of this analysis as a prior, we then analyse colour-dependent clustering, including the red satellite fractions in various luminosity bins as additional free parameters. The latter analysis also simultaneously updates the constraints on the parameters of the luminosity-dependent clustering analysis. 

We find that imposing the extra information of colour-dependent clustering does not change the constraints on the HOD parameters much but improves the overall reduced chi-square significantly due to the use of extra data points ($\chi^2/dof = 142.05/117$ when including colour information compared to  $\chi^2/dof = 65.53/32$ without colour information). In the following, therefore, we will only quote the results of our full analysis which included colour information. Below we will also compare our results with those of a more traditional luminosity dependent analysis which considers luminosity thresholded samples one at a time.

Figure~\ref{fig:corr_all_color_correlation_function} shows the correlation function of galaxies of all colour for the brightest magnitude threshold and the three magnitude bins used in our analysis. The solid circles with error-bars show the measurements from \citep{Zehavi_et_al_2011}; the solid curves show our best-fit model for galaxies of all colour after the colour- dependent clustering information was given. The dotted curve with error-band shows the median and  $\pm 1$-$\sigma$ error from our model. The data points and model curves below the magnitude bin $[-22, -21]$ are separated by $0.25$ dex.  

Figure~\ref{fig:HOD_all_color} shows the corresponding HODs, i.e. population of galaxies of all colour as a function of halo mass in different magnitude thresholds and bins. This is clear from this figure that brighter galaxies tend to live in massive halos as also seen from previous HOD analysis \citep{Zehavi_et_al_2011, Guo_et_al_2015}. 

\subsection{Colour-dependent clustering}
\label{subsec:coldepclust}
Figure~\ref{fig:corr_red_blue} shows the correlation function of red and blue galaxies separately from our analysis. We see that in a given magnitude bin, red galaxies are always more clustered compared to the blue ones. Figure~\ref{fig:HOD_red_blue} shows the population of red and blue galaxies separately as a function of halo mass coming from our analysis in different magnitude bins. 

Figure~\ref{fig:contour_prs_HOD_together} in the Appendix shows the contour plots of the HOD parameters of four magnitude bins discussed above and the satellite red fractions for those four magnitude bins. These $68 \%$ and $95 \%$ contours were obtained by fitting colour-dependent clustering data with parameter priors taken as the posterior from the colour-independent analysis. The marginal distributions of the parameters are shown in the top panels of the figure. 
A detailed list of the constrained values of the satellite red fraction and HOD parameters is shown in Tables~\ref{tab:prs_joint_fit} and~\ref{tab:HOD_global_with_colour}, respectively. 
Figure~\ref{fig:contour_focus} focuses on the joint distributions of parameter pairs which show the largest correlations.

Figures~\ref{fig:log_Mmin_fit} and \ref{fig:alpha} together show our constraints on different HOD parameters with and without colour information and for the analysis of individual magnitude thresholds. 
We see that the constraints on the HOD parameters with or without colour-dependent clustering information agrees well with one another. The constraints due to the analysis of individual magnitude thresholds match with the global analysis except for the two parameters $\sigma_{\log M}$ and $\log M_0$. 
These differences are very likely due to the coupling between parameters for different thresholds introduced by our global likelihood calculation. The fact that these are especially pronounced for $-21\lesssim M_r\lesssim-20$ could be due to the fact that the total number of galaxies in this bin is substantially smaller than in neighbouring bins due to our choice of redshift ranges (see Figure~\ref{fig:SDSS_sample_threshold_bin}).

The right panel of Figure~\ref{fig:alpha} shows the red fraction of the satellite galaxies as a function of magnitudes both from our model and other calibrations. 
We find that the resulting luminosity dependence of the satellite red fraction is significantly shallower than corresponding measurements from the galaxy group catalogue of \citet{Yang_et_al_2007}. In part, this could be due to the known systematic errors in the group finding algorithm, which are at the level of $10$-$15\%$ \citep{campbell+15}; our clustering-based determination is free from such systematics. However, considering the importance of this variable for galaxy evolution studies \citep[see, e.g.,][]{vdb+08}, it will still be interesting to understand these differences in greater detail. We leave this exercise for future work.

\begin{table*}
\begin{tabular}{c|l|l}
Parameter & Parameter values & Covariance of Paarameters $\times 10^{-2}$ \\
\hline
$\log M_{\rm min}$ & [$a_0$, $a_1$, $a_2$] = [$12.33 \pm 0.04$, $-0.85 \pm 0.04$, $0.19 \pm 0.04$] &  cov($a_0$, $a_1$, $a_2$) = $\begin{pmatrix}
0.16 & -0.04 & -0.1 \\
-0.04 &  0.13 &  0.07 \\
-0.1 & 0.07 &  0.13
\end{pmatrix}
$
\\
\hline 
$\sigma_{\log M}$ & [$a_0$, $a_1$, $a_2$] = [$0.44 \pm 0.07$, $-0.16 \pm 0.09$, $0.3 \pm 1.48$] & cov($a_0$, $a_1$, $a_2$) = $\begin{pmatrix}
0.5 & 0.4 & -1.7 \\
0.4 & 0.8 & -8.3 \\
-1.7 & -8.3 & 217.7
\end{pmatrix}$ \\
\hline
$\log M_0$ & [$a_0$, $a_1$] = [$12.24 \pm 0.21$, $-0.54 \pm 0.26$] & cov($a_0$, $a_1$) = $\begin{pmatrix}
4.61 & -4.00 \\
-4.00 &  6.79
\end{pmatrix}
$ \\
\hline
$\log M_1^{\prime}$ & [$a_0$, $a_1$, $a_2$] = [$13.52 \pm 0.02$, $-0.72 \pm 0.02$,  $0.16 \pm 0.02$] & cov($a_0$, $a_1$, $a_2$) = $
\begin{pmatrix}
0.06 & 0.01 & -0.03 \\
0.01 & 0.02 & -0.01 \\
 -0.03 & -0.01 & 0.03
\end{pmatrix}
$\\
\hline
$\alpha$ & [$a_0$, $a_1$, $a_2$] = [ $1.16 \pm 0.02$, $-0.20 \pm 0.04$, $0.10 \pm 0.03$] & cov($a_0$, $a_1$, $a_2$) = $
\begin{pmatrix}
 0.06 & 0.03 & -0.05 \\
 0.03 & 0.17 & -0.12 \\
 -0.05 & -0.12 & 0.11
\end{pmatrix}
$ \\
\hline
$p_{\rm rs}$ & [$a_0$, $a_1$, $a_2$] = [$0.773 \pm 0.008$, $-0.065 \pm 0.010$,  $0.008 \pm 0.009$] & \rm cov ($a_0$, $a_1$, $a_2$) = $\begin{pmatrix}
0.007 &  0.002 & -0.004 \\
0.002 &  0.010 & -0.006 \\
-0.004 & -0.006 &  0.008
\end{pmatrix} $ \\
\hline
\end{tabular}
\caption{Fitting functions of the five HOD parameters and satellite red fraction obtained with global analysis using information of colour-dependent clustering. The first column denotes the parameter concerned, the second column shows the best-fit parameters of the fitting function with diagonal errors on them while the third column shows the full covariance matrix of the parameters. Defining $x \equiv M_r + 20.5$, for the parameter $\sigma_{\log M}$ we have used the form $a_0 + a_1 \erf{(x/a_2)}$ where erf is the error function, while for all other parameters we used the form $a_0 + a_1 x + \ldots a_n x^n$.} \label{tab:HOD_fit}
\end{table*}

\subsection{Fitting functions and applications}
\label{sec:mock_gal_cat}
One of the useful applications of the HOD approach is generating mock galaxy catalogues. To generate mock catalogues of galaxies as a function of magnitudes, one needs to make smooth fitting formulae for the HOD parameters as a function of magnitude \citep{ss09}. 

We prescribe different types of smooth fitting functions to the set of HOD parameters coming from the colour-dependent global analysis. The analytical form of the fitting functions and the best-fit values and uncertainties of the corresponding parameters are shown in Table~\ref{tab:HOD_fit}. Figures~\ref{fig:log_Mmin_fit} and \ref{fig:alpha} show the comparison of our fitting functions with the measured parameter values.

The \emph{left panel} of Figure~\ref{fig:corr_fit_together_check} shows the correlation function of all the galaxies for different magnitude thresholds. The solid points with error bars show the measurements from SDSS data taken from \citet{Zehavi_et_al_2011}. The solid curves show the 2pcf calculated using the fitting functions of the HODs from Table~\ref{tab:HOD_fit} combined with the simulation-based tabulated halo correlation functions. The \emph{right panel} of the Figure shows the cumulative luminosity function as a function of luminosity. We see that our fitting functions, when combined with the simulation-based tabulated theoretical 2pcf, accurately describe the 2pcf and luminosity function of galaxies at all available magnitude thresholds. 

Mock catalogues based on these fitting functions could be useful, e.g., in setting up null hypotheses for testing a variety of `beyond halo mass' effects in galaxy evolution studies, such as assessing the magnitude of assembly bias \citep{Zentner_et_al_2014, Zentner_et_al_2019} and conformity effects at large scales \citep{pkhp15}, the role of the cosmic tidal environment in determining galaxy properties \citep{phs18b,azpm18}, etc. Our parameter constraints are consistent to those obtained with the decorated HOD in \citet{Zentner_et_al_2019}.

For example, \citet{phs18b} showed that a `halo mass only' flavour of HOD parametrisation was sufficient to qualitatively describe the dependence of galaxy clustering in the SDSS on the cosmic tidal environment, but also found some intriguing quantitative differences between the mocks and data in the most anisotropic environments. Since the mocks used by those authors were based on less accurate HOD interpolations than ours, it will be very interesting to revisit that analysis with the more accurate mock catalogues that our fitting functions would allow for.

Additionally, our fitting functions for the HOD parameters can also be directly used in analyses which combine optical properties of galaxies with other properties such as neutral hydrogen (HI) mass. Recently, \citet{Paul_et_al_2018} presented an analysis of galaxy clustering as a function of HI mass by using optical HODs calibrated on SDSS \citep[the same set used by][]{phs18b} with an assumed optical-HI scaling relation which was constrained using MCMC techniques applied to clustering data from the ALFALFA survey \citep{Guo_et_al_2017}. Our HOD fitting functions would be useful in making the resulting best-fit scaling relations more robust and accurate.

\section{Conclusion}
\label{sec:conclude}
In this work, we have revisited the Halo Occupation Distribution (HOD) method of describing the luminosity- and colour-dependence of clustering of SDSS galaxies with the goal of obtaining accurate and self-consistent HOD prescriptions parametrised using convenient fitting functions. Our analysis has combined two techniques that have allowed us to minimise systematic modelling uncertainties while maximising the information content available in the data for parameter estimation.

Firstly, we have calibrated the HOD parameters by using direct measurements of halo correlation functions from $N$-body simulations (section~\ref{sec:sim_stats}). This way of modelling the HOD accurately accounts for the scale-dependence and nonlinearity of halo bias, as well as potential departures of the halo density profile from the universal NFW form. We have also accounted for halo exclusion by modelling halos as hard spheres. Our measurements rely on a suite of dark matter simulations which span the entire range of halo mass and spatial separation required for modelling SDSS projected clustering. We have used multiple independent realisations of our simulation boxes to obtain reliable estimates of the halo correlation functions.

Secondly, we have used a novel global analysis of SDSS clustering measurements. By modelling the projected clustering in a given luminosity bin using HOD parameters for a pair of luminosity thresholds, we were able to self-consistently combine measurements from a range of luminosity bins into a single, global likelihood which simultaneously constrains the HOD parameters for all luminosity thresholds of interest (section~\ref{sec:global_HOD}). This global analysis thus uses all the available information from the data while correctly accounting for correlations between parameter estimates.

In addition to luminosity dependence, we have also included the colour dependence of clustering in our analysis. This substantially improves the quality of our fit (Table~\ref{tab:HOD_global_with_colour}) while also allowing us to place essentially model-independent constraints on the red fraction of satellites $p_{\rm rs}$ (section~\ref{subsec:coldepclust}, Table~\ref{tab:prs_joint_fit}). The resulting luminosity dependence of $p_{\rm rs}$ (right panel of Figure~\ref{fig:alpha}) shows interesting differences from previous calibrations from analyses of galaxy group catalogues, which we will follow up in future work.

\begin{figure*}
\includegraphics[width=0.8\textwidth]{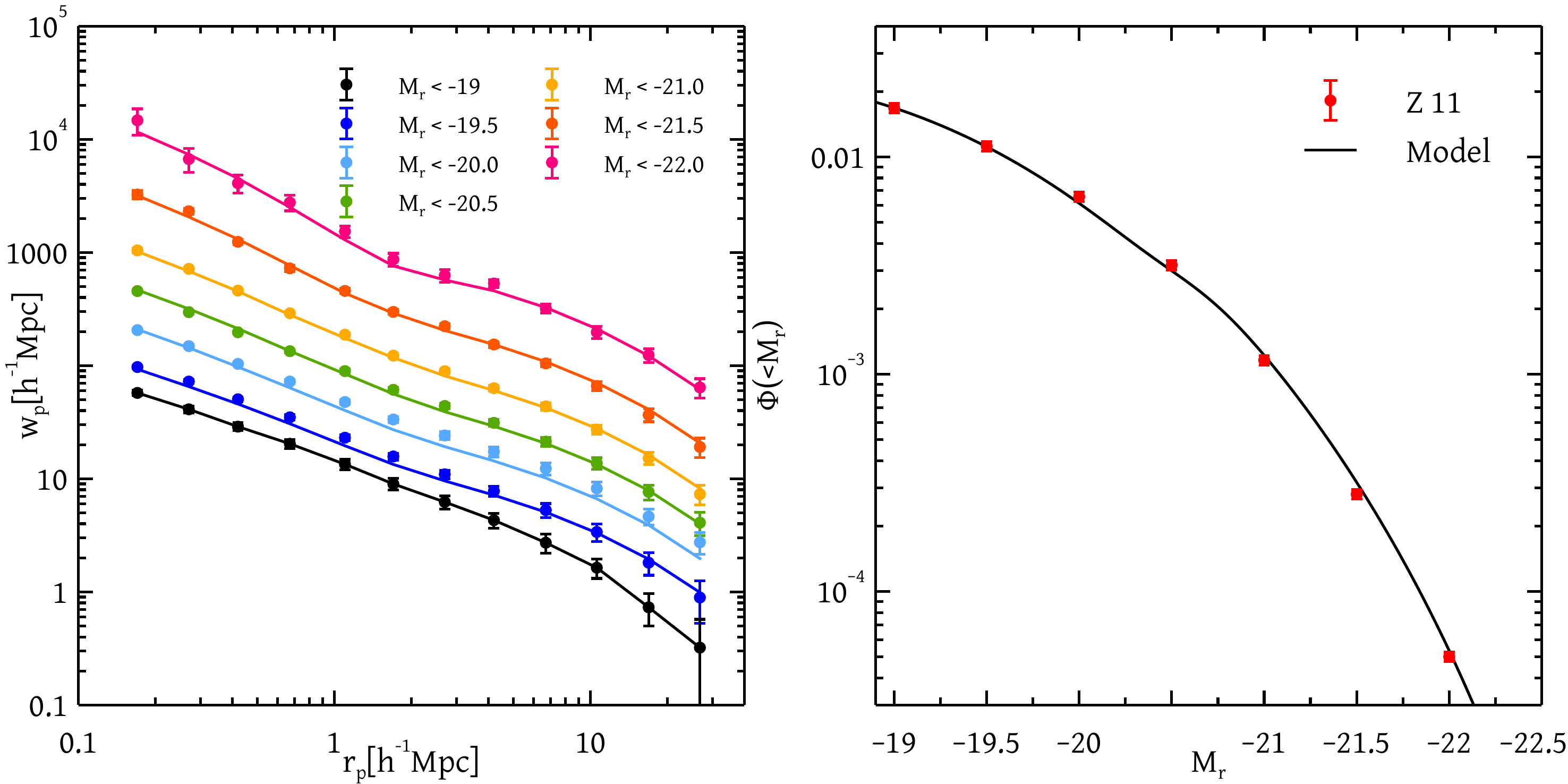} 
\caption{Comparison of clustering and abundance computed using our HOD fitting functions with measurements. \emph{(Left panel:)} Projected correlation function for different magnitude thresholds. Solid circles with error bars show measurements from \citet{Zehavi_et_al_2011}. Solid curves show the correlation function computed using fitting form of the HOD parameters as mentioned in Table~\ref{tab:HOD_fit} combined with our simulation-based theoretical model of halo clustering. (For clarity, we have introduced a $0.25$dex staggering in the measurements and models for all thresholds other than $M_r<-20.5$.)  \emph{(Right panel:)} Cumulative luminosity function as a function of magnitude. Black solid line shows the curve obtained from our theoretical model and red circles show the measurements from \citet{Zehavi_et_al_2011}.} \label{fig:corr_fit_together_check}
\end{figure*}

We have produced simple and accurate fitting functions for the luminosity dependence of all HOD parameters, as well as the satellite red fraction (Table~\ref{tab:HOD_fit}). We have demonstrated that these correctly describe the clustering of samples defined using thresholds which were not included in the analysis (Figure~\ref{fig:corr_fit_together_check}). We expect these fitting functions to be very useful in making mock catalogues of galaxies at low redshift.

Our technique can be easily extended to include halo properties other than halo mass (such as concentration or asphericity) when constructing our tabulated theoretical 2pcfs. The resulting models can then serve, e.g., to calibrate the level of galaxy assembly bias or galactic conformity in the clustering and abundance data. Our mass-only calibrations can themselves also be improved further by considering the possible deviation of the spatial distribution of the satellites from the underlying dark matter distributions, as well as allowing for colour-dependent differences in these distributions. Overall, we do not find strong correlations between the parameters of different magnitude thresholds (Figure~\ref{fig:contour_prs_HOD_together}, see also Figure~\ref{fig:contour_focus}). The constraints on HOD parameters can be further improved if one uses the measurements of anisotropic galaxy correlations \citep{Guo_et_al_2015}. Incorporating these improvements is the subject of work in progress.

\section*{Acknowledgments}
We thank R. Srianand, K. Subramanian and T. R. Choudhury for useful discussions. This work has used the open source computing packages \textsc{NumPy}  \citep[][http://www.numpy.org]{vanderwalt-numpy}, \textsc{SciPy} \citep{scipy_package} and the plotting softwares \textsc{Veusz} (https://veusz.github.io/) and corner.py \citep{corner}. 
The research of AP is supported by the Associateship Scheme of ICTP, Trieste and the Ramanujan Fellowship awarded by the Department of Science and Technology, Government of India. NP acknowledges the financial support from the Council of Scientific and Industrial Research (CSIR), India as a Shyama Prasad Mukherjee  Fellow. IP acknowledges the hospitality and facilities provided by the Korea Institute for Advanced Study and School of Liberal Arts, Seoul-Tech, South Korea where part of this work was completed. We gratefully acknowledge the use of high performance computing facilities at IUCAA, Pune. We sincerely thank the anonymous referee for the valuable comments which led to the significant improvement of the draft. 

\bibliography{HOD_calib_bibliography}

\appendix
%
\section{Simulation-based halo model} \label{sec:2PCF_calculation_formulae}
In this section, we discuss how to compute the 2pcf of galaxies using measurements of different quantities from sinulations. In the first subsection, we describe how to compute the 2pcf and abundance for magnitude thresholds, then in the next subsection, we discuss how to compute those quantites in binned measurements using the thresholded HOD. Finally, in the next subsection, we discuss how to compute the 2pcf for red and blue galaxies.
\subsection{Computation of abundance and clustering for individual magnitude threshold}
If we have a total of $N$ bins of halo mass and in a given halo mass bin $\log m_i \pm \der \log m_i /2$, we have HODs $\fcen(m_i)$ and $\Ns(m_i)$ cf. equation~\eqref{eq:fcen}, \eqref{eq:Ns}, then  the average comoving number density of galaxies can be written as,
\begin{align}
\bar{n}_g = \sum_{i = 1}^N \left[ \fcen(m_i) + \Nscript (m_i) \right] n(m_i) \der \log m_i \,\, . \label{eq:avg_gal_num_density}
\end{align}   
It is worth to note that all the HOD quantities and the ones derived from it depends on the properties of the galaxies, in this case luminosity $L$, which we have not written explicitly. In equation~\eqref{eq:avg_gal_num_density}, $n(m_i)$ is the comoving number-density of halos per unit logarithmic mass in the halo mass-range $\vert \log (m/m_i) \vert \leq \der \log m_i /2$ and $\Nscript (m_i) = \fcen(m_i) \Ns (m_i)$. If the density profile of the galaxies inside dark matter halos in the $i$-th mass bin be $\rho(r|m_i)$ and the convolution of the density profile with itself be $\lambda (r|m_i)$, then the $1$-halo correlation function of the galaxies can be written as,
\begin{align}
\xi_{gg}^{1h} (r) &= \sum_{i = 1}^{N} \der \log m_i n(m_i) \frac{\fcen (m_i)}{\bar{n}_g^2} \left[ 2 \Ns(m_i) \frac{\rho(r|m_i)}{m_i} \right. \notag \\
& \left. + \Ns^2(m_i)  \frac{\lambda(r|m_i)}{m_i^2} \right] \,\, . \label{eq:1h}
\end{align}  
The 2-halo term of the correlation function of the galaxies can be similarly computed as, 
\begin{align}
\xi_{gg}^{2h}(r) &= \sum_{i = 1}^{N} \sum_{j = 1}^N \der \log m_i \der \log m_j \frac{n(m_i) n(m_j)}{\bar{n}_g^2} \notag \\
& \times \left[ \fcen(m_i)\fcen(m_j) \xi_{hh}^{cc}(r|m_i, m_j) + 2 \fcen(m_i)  \Nscript (m_j) \right. \notag \\
& \left. \times \xi_{hh}^{cs} (r|m_i, m_j) + \Nscript(m_i) \Nscript(m_j) \xi_{hh}^{ss}(r|m_i, m_j)  \right] \,\, , \label{eq:2h}
\end{align}
where $\xi_{hh}^{cc}$, $\xi_{hh}^{cs}$ and $\xi_{hh}^{ss}$-s are two-point correlations between central-central, central-satellite and satellite-satellite pairs respectively in two different halos. The above equations are just generalisations of the analytical 2pcf formulae in HOD prescriptions \citep{Cooray_Sheth_2002, zg16}. Once we have these two terms we can compute total correlation function $\xi(r) = \xi^{1 h}(r) + \xi^{2h} (r)$ and then compute the projected 2pcf $w_p$ using  \citep{Davis_Peebles_1983},
\begin{align}
w_p(r_p) &= 2 \int_{r_p}^{\infty} \frac{\der r \xi (r)}{\sqrt{r^2- r_p^2}} \,\, . \label{eq:wp_peebles}
\end{align}
The upper limit in the above integration is practically not $\infty$, rather it is $r_{\rm max} = \sqrt{r_p^2 + \pi_{\rm max}^2}$, where the two quantities $r_p$ and $\pi$ are the separation between two galaxies perpendicular and parallel to the line of sight respectively. 

Our procedure of convolving the HOD-s wit halo statistics is a robust and fast technique. But it has its own drawback. We have assumed that the central galaxy always live at the center of the halos and the density profile of the galaxies trace the underlying density profile of dark matter halos. These assumptions are not always true and can have sub-percent level effect at small scale correlation functions. An alternative approach where one obtain the galaxy population by directly populating the halos with HOD-s has the potential to overcome these issues.
\subsection{Computation of abundance and clustering in a magnitude bin}
\label{sec:HOD_fomulae_in_bin}
Now we will see how to compute the correlation and abundance of galaxies in a given magnitude bin using the HODs of two adjacent magnitude thresholds. Let's denote by $\fcen(L_{12}|m)$ the fraction of $m$-halos having a central galaxy with luminosity in the range $L_1<L<L_2$. We also define $\Ns(L_{12}|m)$ to be the average number of satellite galaxies of luminosity $L_1<L<L_2$ residing in $m$-halos which have a central galaxy with luminosity $L>L_1$. Then the binned-HODs can be computed in the following way \citep{pkhp15, pp17b}, 
\begin{align}
\fcen(L_{12}|m) &= \fcen(>L_1|m) - \fcen(>L_2|m) \label{eq:binned_fcen_from_threshold} \,\, \\
\Nscript(L_{12}|m) &= \Nscript(>L_1|m) - \Nscript(>L_2|m) \,\, , \\
\Ns(L_{12}|m) &= \Nscript(L_{12}|m)/\fcen(>L_1|m) \,\, . \label{eq:binned_Ns_from_threshold}
\end{align}
Having defined these binned HODs, the calculation of  the abundance and the 2-halo term of the correlation function is straighforward. One just needs to replace $\fcen(>L|m)$ and $\Nscript(>L|m)$ with $\fcen(L_{12}|m)$ and $\Nscript(L_{12}|m)$ in equation~\eqref{eq:avg_gal_num_density} and \eqref{eq:2h}. But for the 1-halo correlation function, we need to keep in mind that the first term of the correlation function will be the correlation between a central galaxy of luminosity $L_1<L<L_2$ and satellites of luminosity $L_1<L<L_2$ residing in a $m$-halo having a central of luminosity $L>L_1$. The second term will be the correlation between satellites of luminosity $L_1<L<L_2$ living inside a halo having a central galaxy of luminosity $L>L_1$. Therefore, then modified 1-halo correlation function will be,
\begin{align}
\xi_{gg}^{1h}(L_{12}|r) &= \sum_{i = 1}^{N} \der \log m_i \frac{n(m_i)}{\bar{n}_g^2 \fcen(>L1)} \left[ 2 \fcen(L_{12}|m_i) \right. \notag \\
& \left. \times \Nscript(L_{12}|m_i) \frac{\rho(r|m_i)}{m_i} + \Nscript^2(L_{12}|m) \frac{\lambda(r|m_i)}{m_i^2} \right] \,\, . \label{ref:xi_1h_lum_bin}
\end{align} 
\subsection{Computation of red and blue clustering in a given magnitude bin}
\label{sec:colour_2PCF_formulae}
Once we know how to compute the 2pcf of galaxies in a given magnitude bin, we can proceed to model the colour-dependent clustering.  We need the following red and blue HODs to compute the colour-dependent correlation function, 
\begin{align}
\phi_g({\rm red}|M_r^{\rm bin}, m) &=  p({\rm red}|M_r^{\rm bin}, g)\phi_g(M_r^{\rm bin}|m) \,\,\,\, \rm and \notag \\
\phi_g({\rm blue}|M_r^{\rm bin}, m) &= p({\rm blue}|M_r^{\rm bin}, g)\phi_g(M_r^{\rm bin}|m) \,\,. \label{eq:red_blue_HOD}
\end{align}
Here $g$ is the galaxy type, either central or satellite.  In the above equations we have assumed that red (or blue) fractions of the galaxies are independent of their halo mass. In the following, let's derive the working equations for the red fraction first. The corresponding equations for blue galaxies will follow. In our model, we will keep the red fraction of satellite galaxies, $p({\rm red}|M_r^{\rm bin}, {\rm sat}) = p_{\rm rs}$ as a free parameter. Then the red fraction of the centrals can be computed very easily in the following way,
\begin{align}
p({\rm red}|M_r^{\rm bin}, {\rm cen}) &= \frac{\left[ p({\rm red}|M_r^{\rm bin}) - p_{\rm rs} \times \bar{p}({\rm sat}|M_r^{\rm bin}) \right]}{\bar{p}({\rm cen}|M_r^{\rm bin})} \,\, .
\end{align}
In the previous expression, $\bar{p}({\rm cen (sat)}|M_r^{\rm bin})$ is the fraction of galaxies which are  central (satellite) in a given magnitude bin. This can be computed from the HOD in the following way,
\begin{align}
\bar{p}({\rm cen (sat)}|M_r^{\rm bin}) = \frac{\int_{M_r^{\rm max}}^{M_r^{\rm min}} \der M_r \int \der m \phi_{{\rm cen(sat)}}(M_r|m) n(m)}{\int_{M_r^{\rm max}}^{M_r^{\rm min}} \der M_r \int \der m \phi(M_r|m) n(m)} \,\, .
\end{align}
We are now only left with the quantity $p({\rm red}|M_r^{\rm bin})$, i.e. the red fraction of all galaxies in a given magnitude bin. This we can get directly from SDSS data using \citet{Zehavi_et_al_2011} definition of red and blue galaxies. For the blue galaxies, we can write,
\begin{align}
p({\rm blue}|M_r^{\rm bin}, {\rm sat}) = 1 - p({\rm red}|M_r^{\rm bin}, {\rm sat}) \notag \\
p({\rm blue}|M_r^{\rm bin}, { \rm cen}) = 1 - p({\rm red}|M_r^{\rm bin}, {\rm cen}) \,\,.
\end{align}
Thus having obtained the HODs for red and blue galaxies separately we can use the formalism defined  in the previous subsection to compute the correlation function of red and blue galaxies in a given magnitude bin. 

\begin{figure*}
\includegraphics[scale=0.1]{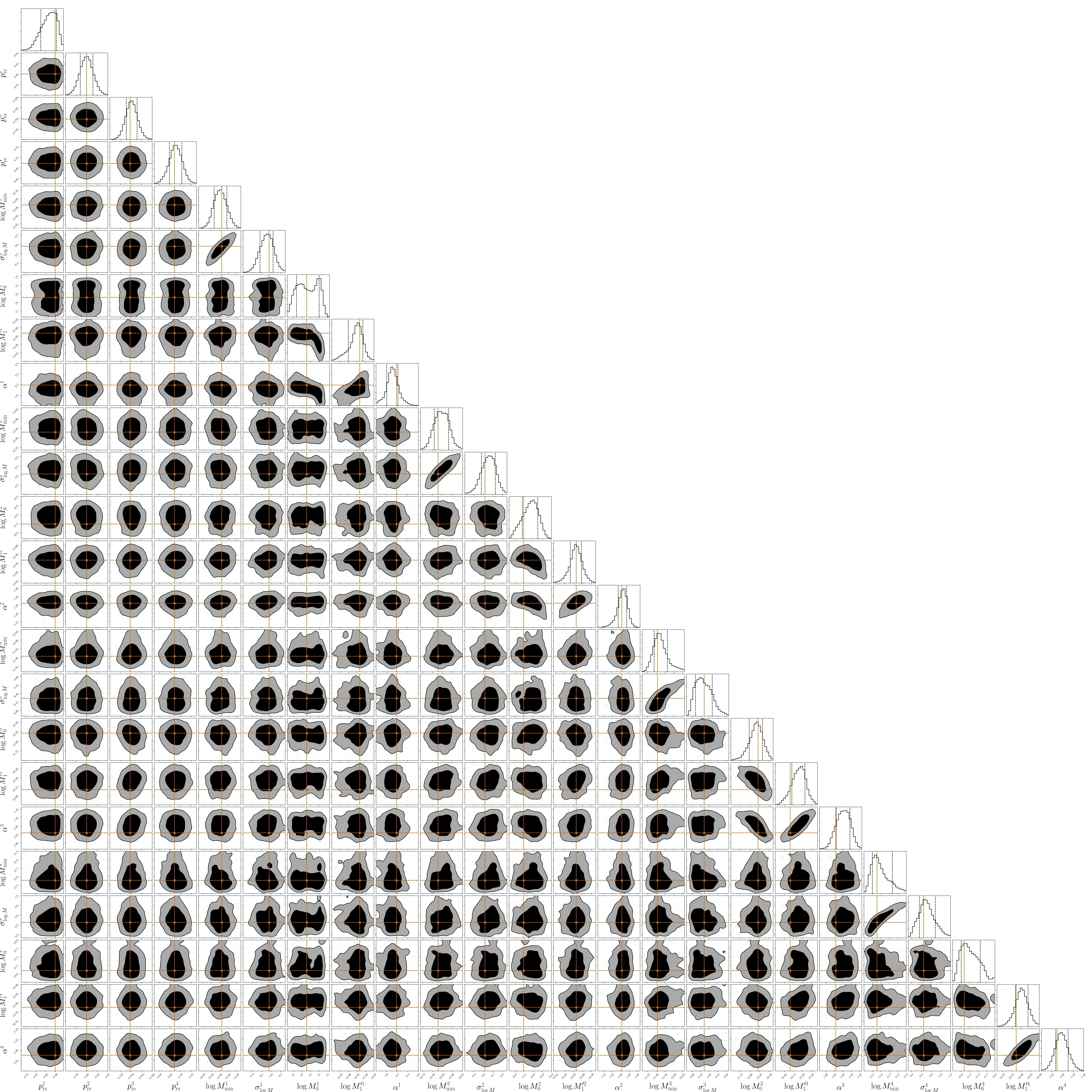} 
\caption{Joint contours of all the HOD parameters and the parameters for the satellite red fraction. We have shown $68^{\rm th}$ and $95^{\rm th}$ percentile contours in all the panels. For the satellite red fraction parameters $p_{\rm rs}$, the superscript numeric symbols correspond to magnitude bins [(-23,-22), (-22, -21), (-21, -20), (-20, -19)] respectively and the numeric subscripts in the HOD parameters correspond to magnitude thresholds [-22, -21, -20, -19] respectively. The vertical and horizontal orange lines denote the best-fit values of our parameters. } \label{fig:contour_prs_HOD_together}
\end{figure*} 

\begin{figure}
\includegraphics[scale=0.24]{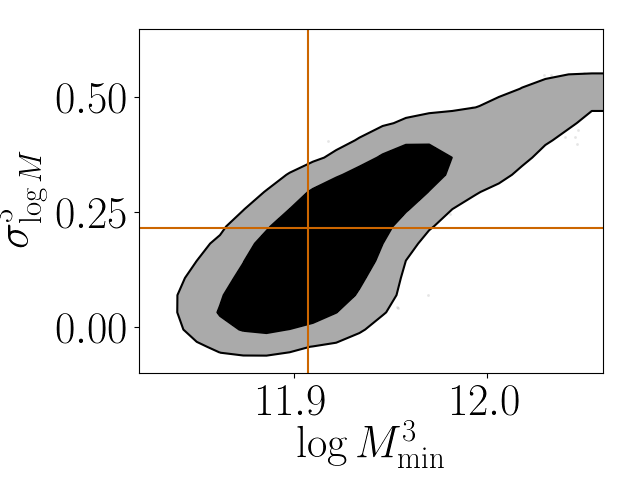} \,\,\,\,
\includegraphics[scale=0.24]{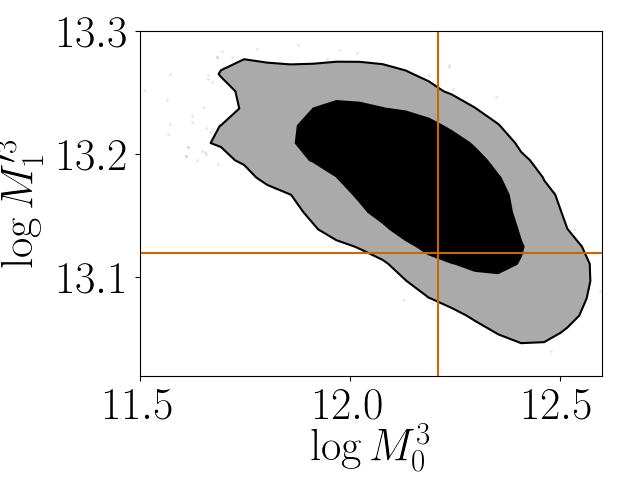} 
\\
\includegraphics[scale=0.25]{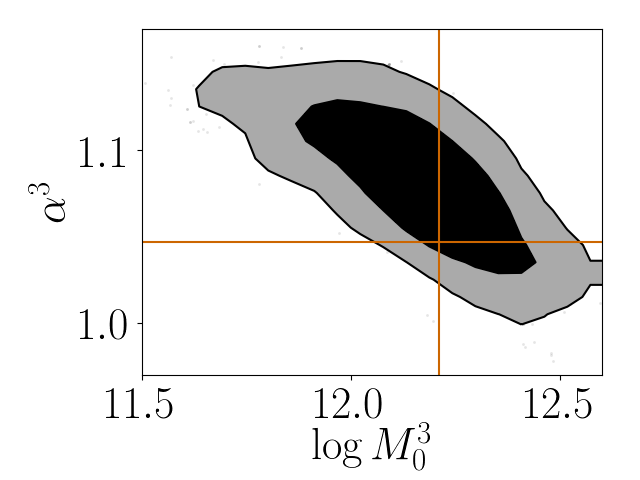} \,\,
\includegraphics[scale=0.25]{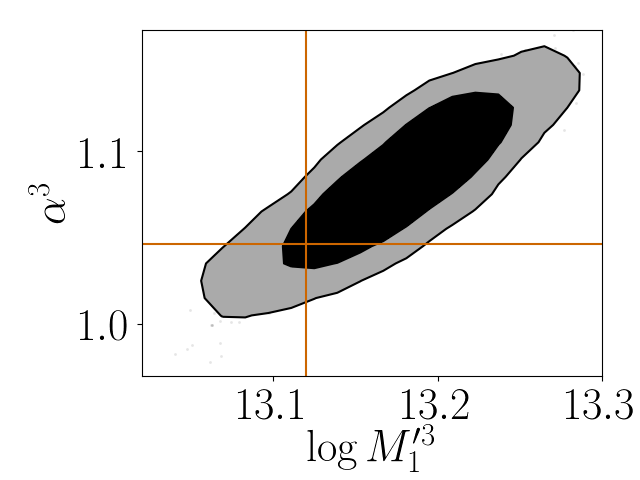} 
\caption{Contour plots for a few pairs of parameters showing the strongest correlations for the magnitude threshold $M_r < -20.0$ taken from Figure~\ref{fig:contour_prs_HOD_together}. The correlation between the corresponding parameters for other luminosity thresholds are similar. We did not find any significant correlation between parameters across different luminosity thresholds.}
\label{fig:contour_focus}
\end{figure}

\section{Weighting Scheme for simulation measurements and their errors} \label{sec:simulation_weight}
If a given quantity $x$ is measured from different realisations as a set $\{x_i\}$ with corresponding uncertainties $\{\sigma_i\}$, then the weighted average which maximises the joint likelihood function over all realisations has the following expression,
\begin{align}
\bar{x} &= \frac{\sum_i x_i /\sigma_i^2}{\sum_i 1/\sigma_i^2} \,\, . \label{eq:x_avg}
\end{align}  
For our work, we assume a Poisson distribution of the quantities, so the relative error on the mean value of the measured quantities in each realisation and in each halo mass bin scales as the inverse of the square root of the number of particles associated with each component measured. So with the approximation that the mean of the quantities measured in each realisation and in each halo mass bin is same, we can define the unnormalized weights to be equal to the number of particles in that halo mass bin and in that realisation which has been used to measure the quantity. It is important to note that throughout this discussion, by `realisation', we mean all available realisations from all the boxes. 

Therefore, for the halo mass function $n(m)$, weight is $w_i = N_i^h(m)/\sum_i N_i^h(m)$, where $N_i^h(m)$ is the number of available halos in the $i-th$ realisation in the halo mass bin $m \pm \der m/2$. For $\rho(r|m)$, the weight will be $w_i  = N_i^h (m) N_i^p (m)/(\sum_i N_i^h(m) N_i^p(m))$. Here $N_i^p(m)$ is the number of available dark matter particles in the $i-th$ realisation and in the halo mass bin $m \pm \der m/2$. Now $N_i^p(m) = m/m_p^i$, where $m_p^i$, the mass of each dark matter particle in the $i-th$ realisation, is computed as $m_p^i = \Omega_m \rho_c V_i/\mathbb{N}_i^p$. In this expression, $\mathbb{N}_i^p$ is the total number of dark-matter particles in the $i-th$ realisation of the box. Similarly, for the quantity $\lambda(r|m)$ the weight will be $N_i^h(m)N_i^p(m)^2$. For simplicity, we assume the weights of the normalized quntities $\rho(r|m)/m$ and $\lambda(r|m)/m^2$ to be same as their unnormalized counterparts. Furthermore, wherever the two quantities $n(m)$ and $\bar{n_g}$ appear, we will take them to be the weighted average over all the realizations. So in case of error estimation, we neglect the errors coming from them. Similarly, the weights for  the 2-halo quantities will be following,
\be
w_i^{cc}(m, m^{\prime}) = \frac{N_i^h(m) N_i^h(m^{\prime})}{\sum_i N_i^h(m) N_i^h(m^{\prime})}
\ee
\be
 w_i^{cs} (m, m^{\prime}) = \frac{N_i^h(m) N_i^h(m^{\prime})N_i^p(m^{\prime})}{\sum_i N_i^h(m) N_i^h(m^{\prime})N_i^p(m^{\prime})}
\ee
\be
w_i^{ss}(m, m^{\prime}) = \frac{N_i^h(m)N_i^p(m) N_i^h(m^{\prime})N_i^p(m^{\prime})}{\sum_i  N_i^h(m)N_i^p(m) N_i^h(m^{\prime})N_i^p(m^{\prime})} \,\, . \label{eq:weight_2h}
\ee
The weights for the term $w_i^{sc}$ is very similar to that of $w_i^{cs}$. Once we know the weights to assign to each of the components, we can find the error associated to their weighted mean. The standard error on the mean has the following expression,
\begin{align}
s^2(m) &= \frac{\sum_i w_i^2}{1-\sum_iw_i^2} \sum_i w_i (x_i - \bar{x})^2 \,\, , \label{ew:weighted_variance}
\end{align}  
where $\bar{x} = \sum_i w_i x_i $.  This is just a schematic equation which will change the exact form depending on whether we are calculating the errors on the 1-halo or 2-halo terms of the correlation function. 

Once we have the mean and error on that mean of the quantities in each halo mass bin, we can compute the mean of the quantites and error on that mean for a given thick halo mass range $m_1<m<m_2$. If the schematic quantity is $x(m)$ with mean $\bar{x}(m)$ and error of the mean $\sigma_{\bar{x}}(m)$ in each halo mass bin, then the weighted average of that quantity in the halo mass range $m_1 < m < m_2$ will be,
\begin{align}
\bar{\bar{x}} \big |_{m_1<m<m_2} &= \frac{\sum_{m_1}^{m_2} \bar{x}(m)/\sigma_{\bar{x}}(m)^2}{\sum_{m_1}^{m_2} 1/\sigma_{\bar{x}}(m)^2} \,\, , \label{eq:mean_combining_bin}
\end{align}
and the error on this mean quantity will be,
\begin{align}
\sigma_{\bar{\bar{x}}}^2 &= \frac{\sum_m 1/\sigma_{\bar{x}}^4(m) \sigma_{\bar{x}}(m)^2}{(\sum_m 1/\sigma_{\bar{x}}^2(m))^2} \notag \\
&= \frac{1}{\sum_m 1/\sigma_{\bar{x}}(m)^2} \label{eq:error_combining_bin}
\end{align}
Formulae from equation~\eqref{eq:x_avg} to \eqref{eq:error_combining_bin} have been used to compute errors in different components of the correlation function in different thick halo mass bins as shown in Figures~\ref{fig:hmf_from_sim} to \ref{fig:xi_cs_off_diag_from_sim}.

\label{lastpage}

\end{document}